\documentclass[journal,onecolumn]{IEEEtran}
\usepackage{amssymb}
\usepackage{graphicx}
\usepackage{epstopdf}
\usepackage[tbtags]{amsmath}
\usepackage{amsfonts}
\usepackage{mathrsfs}
\usepackage{cite,url}
\usepackage{upquote}
\usepackage{pifont}
\usepackage{amsmath}
\newcommand{\ra}{\rightarrow}
\usepackage{units}
\newcommand{\I}{\mathscr{I}}
\newcommand{\Ia}{\mathscr{I}_{\alpha}}
\newcommand{\X}{\mathbb{X}}
\newcommand{\hatt}{\hat{\theta}}

\newcommand{\R}{\mathbb{R}}

\newtheorem{thm}{Theorem}
\newtheorem{theorem}[thm]{Theorem}
\newtheorem{lemma}[thm]{Lemma}
\newtheorem{definition}[thm]{Definition}
\newtheorem{proposition}[thm]{Proposition}
\newtheorem{corollary}[thm]{Corollary}

\newtheorem{example}{Example}

\begin{document}
\title {Minimization Problems Based on\\ Relative $\alpha$-Entropy II: Reverse Projection}
\author{
M.~Ashok~Kumar and Rajesh~Sundaresan
\thanks{M.~Ashok~Kumar was supported by a Council for Scientific and Industrial Research (CSIR) fellowship and by the Department of Science and Technology. R.~Sundaresan was supported in part by the University Grants Commission by Grant Part (2B) UGC-CAS-(Ph.IV) and in part by the Department of Science and Technology. A part of the material in this paper (Section \ref{p2:sec:forwardprojection} alone) was presented at the National Conference on Communication (NCC 2015), Mumbai, India, held during February 2015 \cite{201410NCC_KumSun}.}
\thanks{M.~Ashok~Kumar and R.~Sundaresan are with the ECE Department, Indian Institute of Science, Bangalore 560012, India.}
}

\maketitle

\begin{abstract}
In part I of this two-part work, certain minimization problems based on a parametric family of relative entropies (denoted $\mathscr{I}_{\alpha}$) were studied. Such minimizers were called forward $\mathscr{I}_{\alpha}$-projections. Here, a complementary class of minimization problems leading to the so-called reverse $\mathscr{I}_{\alpha}$-projections are studied. Reverse $\mathscr{I}_{\alpha}$-projections, particularly on log-convex or power-law families, are of interest in robust estimation problems ($\alpha >1$) and in constrained compression settings ($\alpha <1$). Orthogonality of the power-law family with an associated linear family is first established and is then exploited to turn a reverse $\mathscr{I}_{\alpha}$-projection into a forward $\mathscr{I}_{\alpha}$-projection. The transformed problem is a simpler quasiconvex minimization subject to linear constraints.
\end{abstract}


\section{Introduction}
\label{p2:sec:introduction}
This paper is a continuation of our study of minimization problems based on a parametric generalization of relative entropies, denoted $\mathscr{I}_{\alpha}$. See (\ref{p2:alphadiv_expanded}) for the definition of $\mathscr{I}_{\alpha}(P,Q)$, where $P$ and $Q$ are probability measures on an alphabet set $\mathbb{X}$. We say ``parametric generalization of relative entropy'' because $\lim_{\alpha\to 1}\mathscr{I}_{\alpha}(P,Q) = \mathscr{I}(P\|Q)$, the usual relative entropy or Kullback-Leibler divergence. In part I \cite{2014xxManuscript1_KumSun}, we showed how $\mathscr{I}_{\alpha}$ arises and studied the problem of a {\em forward $\mathscr{I}_{\alpha}$-projection}, namely
\[
 \displaystyle\min_{P\in \mathbb{E}}\mathscr{I}_{\alpha}(P,R),
\]
where $R$ is a fixed probability measure on $\mathbb{X}$ and $\mathbb{E}$ is a convex set of probability measures on $\mathbb{X}$. In this paper, we shall study {\em reverse $\mathscr{I}_{\alpha}$-projection}, namely
\[
 \displaystyle\min_{P\in \mathbb{E}}\mathscr{I}_{\alpha}(R,P).
\]
The minimization now is with respect to the second argument of $\mathscr{I}_{\alpha}$. Such problems arise in robust parameter estimation and constrained compression settings. The family $\mathbb{E}$ is usually a parametric family such as the exponential family, or its generalization, called the {\em $\alpha$-power-law family}.

We shall bring to light the geometric relation between the {\em $\alpha$-power-law family} and a {\em linear family}\footnote{Example linear families are (1) the set of probability measures $P$ on $\mathbb{X}$ such that $\sum_x P(x)f(x) = 0$ for some $f\colon\mathbb{X}\to \mathbb{R}$, and (2) finite intersections of such sets. If there is an additive structure on $\mathbb{X}$, a concrete example is the set of all probability measures with a fixed mean.} of probability measures. We shall turn the reverse $\mathscr{I}_{\alpha}$-projection problem on an $\alpha$-power-law family into a forward $\mathscr{I}_{\alpha}$-projection problem on a linear family. The latter turns out to be a minimization of a quasiconvex objective function subject to linear constraints.

The outline of the paper is as follows. In Section \ref{p2:sec:preamble}, we motivate reverse $\mathscr{I}_{\alpha}$-projections for the cases $\alpha>1$ and $\alpha<1$. In Section \ref{p2:sec:setting}, we define the required terminologies and highlight the contributions of the paper. In Section \ref{p2:sec:reverse_projection}, we study the existence of a reverse $\mathscr{I}_{\alpha}$-projection on general log-convex sets. In Section \ref{p2:sec:forwardprojection}, we provide simplified proofs of some essential results from \cite{2014xxManuscript1_KumSun} on the forward $\mathscr{I}_{\alpha}$-projection. Our simplified proofs also serve the purpose of keeping this paper self-contained. In Section \ref{p2:sec:orthogonality}, we explore the geometric relation between the $\alpha$-power-law and the linear families, and then exploit it to study reverse $\mathscr{I}_{\alpha}$-projection on $\alpha$-power-law families. The paper ends with some concluding remarks in Section \ref{p2:sec:concludingRemarks}.

\section{Motivations}
\label{p2:sec:preamble}
The purpose of this section is to motivate reverse $\mathscr{I}_{\alpha}$-projections. The motivation for $\alpha >1$ comes from robust statistics. The motivation for $\alpha <1$ comes from information theory as well as from a strong similarity of the outcomes with the $\alpha =1$ (relative entropy) setting.
\subsection{Reverse $\I$-projection}
\label{p2:subsec:reverse-I}

Let $\X$ be a finite alphabet set and let $\mathbb{E} = \{ P_{\theta} \colon \theta \in \Theta \}$ denote a family of probability measures on $\X$ indexed by the elements of the index set $\Theta \subset \R^k$ for some $k$. Let $x_1, x_2, \ldots, x_n$ be $n$ samples drawn independently and with replacement from $\X$ according to an unknown probability measure $P_{\theta}$ belonging to $\mathbb{E}$. The maximum likelihood estimate (MLE) of $\theta$, denoted $\hatt$, is the element of the index set $\Theta$ that maximizes the likelihood (if it exists), i.e.,
\begin{equation}
  \label{p2:eqn:theta-ml}
  \hatt = \arg \max_{\theta \in \Theta} ~ \prod_{i=1}^n P_{\theta}(x_i).
\end{equation}
Let $\hat{P}$ denote the empirical measure of the $n$ samples $x_1, \ldots, x_n$, i.e.,
\[
  \hat{P} := \frac{1}{n} \sum\limits_{i=1}^n \delta_{x_i},
\]
where $\delta_a$ denotes the Dirac mass at $a$. One may then write
\begin{eqnarray*}
  \frac{\prod_{i=1}^n P_{\theta}(x_i)}{\prod_{i=1}^n \hat{P}(x_i)} & = & \prod_{i=1}^n \frac{P_{\theta}(x_i)}{\hat{P}(x_i)} \\
    & = & \prod_{x \in \X} \left(\frac{P_{\theta}(x)}{\hat{P}(x)}\right)^{n \hat{P}(x)}\\
    & = & \exp \{ -n \I(\hat{P}\|P_{\theta}) \},
\end{eqnarray*}
where
\[
  \I(P \| Q) := \sum\limits_{x \in \X} P(x) \log \frac{P(x)}{Q(x)}
\]
is the relative entropy\footnote{The usual convention is $p \log \frac{p}{q} = 0$ if $p = 0$ and $+\infty$ if $p > q = 0$.} of $P$ with respect to $Q$. Hence the MLE is the minimizer (if it exists)
\begin{equation}
  \label{p2:eqn:theta-ml-log}
  \hatt = \arg \min_{\theta \in \Theta} \I(\hat{P} \| P_{\theta}),
\end{equation}
and the corresponding probability measure $P_{\hatt}$ is known as the \emph{reverse $\I$-projection of $\hat{P}$ on the family $\mathbb{E}$}. Such reverse projections, particularly those related to robustifications of the MLE, are the subject matter of this paper.

Observe that the MLE depends on the samples only through their empirical measure. Let us write the MLE as a function of the empirical measure in a different way. Assume that the family $\mathbb{E}$ is sufficiently smooth in the parameter $\theta$ on account of which we can define the {\em score function} as $s(\cdot ~;\theta) := \nabla_{\theta}\log P_{\theta}(\cdot)$, the gradient of $\log P_{\theta}(\cdot)$ with respect to $\theta$. The first order optimality criterion applied to (\ref{p2:eqn:theta-ml}) after taking logarithms yields the so-called estimating equation for the MLE:
\begin{eqnarray*}
 \frac{1}{n} \sum\limits_{i=1}^n s(x_i;\theta) = 0;
\end{eqnarray*}
the MLE $\hatt$ solves this equation. Write $E_P[\cdots]$ for expectation with respect to $P$. Noting that the score function satisfies
\[
  E_{P_{\theta}}[s(X; \theta)] = 0 \quad \forall P_{\theta},
\]
the estimating equation for the MLE can be rewritten as
\begin{equation}
 \label{p2:eqn:likelihood-equation-i}
 \frac{1}{n} \sum\limits_{i=1}^n s(x_i;\theta) = E_{P_{\theta}}[s(X; \theta)],
\end{equation}
which is the same as
\begin{equation}
  \label{p2:eqn:likelihood-equation}
  E_{\hat{P}} [s(X;\theta)] = E_{P_{\theta}}[s(X; \theta)].
\end{equation}
If we write $T(\hat{P})$ for the $\theta$ that solves (\ref{p2:eqn:likelihood-equation}), we then have $\hatt = T(\hat{P})$. The estimator $T(\hat{P})$ is {\em Fisher consistent}\footnote{An estimator that maps an empirical measure to an element in $\Theta$ is {\em Fisher consistent} if it is continuous and maps $P_{\theta}$ to the true parameter $\theta$. See \cite[Sec.~5c.1]{1973Rao_LSIIA}}, a fact that can be easily checked using (\ref{p2:eqn:likelihood-equation}).

\subsection{Reverse $\Ia$-projection: $\alpha > 1$}
\label{p2:subsec:reverse-I-alpha_g1}

Though the MLE is known to possess many good properties, asymptotic efficiency being an example, it is not appropriate when some of the data entries ($x_i$) are contaminated by outliers. To achieve robustness, one may consider scaling the scores $s(x_i; \theta)$ in the left-hand side of (\ref{p2:eqn:likelihood-equation-i}) by weights $w(x_i;\theta)$ that weigh down outlying observations ``relative to the model'' (see for example Basu et al. \cite{1998xxBIO_Bas}). This type of robustification, along with the requirement of Fisher consistency, is accomplished by the estimator that maps the empirical measure $\hat{P}$ to the $\theta$ that solves the equation
\begin{eqnarray}
\label{p2:eqn:nonnormalized_estimating_equation}
 E_{\hat{P}}[w(X;\theta) s(X;\theta)] = E_{P_{\theta}}[w(X;\theta) s(X;\theta)].
\end{eqnarray}
Basu et al. \cite{1998xxBIO_Bas} proposed the natural weighting $w(x;\theta) = P_{\theta}(x)^c$ where $c>0$.  As another robustification procedure, Basu et al. \cite{1998xxBIO_Bas} proposed a weighting of the model {\em by itself}, motivated by the works of Field and Smith \cite{1994xxISR_Fie_Smi} and Windham \cite{1995xxJRSS_Win}, prior to solving the estimating equation. Their procedure is as follows. Given a measure $Q$, its weighting with respect to a parameter $c > 0$ and a model $\theta \in \Theta$, denoted $Q^{(c, \theta)}$, is given by
\[
  Q^{(c, \theta)}(x) = \frac{w(x;\theta) Q(x)}{\sum\limits_{y \in \X} w(y;\theta) Q(y)}, \quad x \in \X,
\]
where the dependence on $c$ is through the weighting $w(x;\theta) = P_{\theta}(x)^c$ as before. Observe that $(P_{\theta})^{(c,\theta)}$ weighs $P_{\theta}$ by itself, namely the weighting parameters are $c$ and $\theta$, and $(P_{\theta})^{(c,\theta)}$ is the probability measure proportional to ${P_{\theta}}^{c+1}$. The Basu et al. procedure\footnote{This procedure may be viewed as a generalization of the self-weighting procedure suggested by Windham \cite[p.~604]{1995xxJRSS_Win}.} \cite{1998xxBIO_Bas} is to find the $\theta$ that solves the equation
\begin{eqnarray}
\label{p2:eqn:normalized_estimating_equation}
 E_{(\hat{P})^{(c, \theta)}}[s(X;\theta)] = E_{(P_{\theta})^{(c,\theta)}}[s(X;\theta)];
\end{eqnarray}
the $\hat{P}$ and $P_{\theta}$ of (\ref{p2:eqn:likelihood-equation}) are replaced by the model reweighted $(\hat{P})^{(c, \theta)}$ and $(P_{\theta})^{(c, \theta)}$, respectively. It is clear that the corresponding estimator is Fisher consistent. Now (\ref{p2:eqn:normalized_estimating_equation}) can be rewritten as
\begin{eqnarray*}
 \frac{\frac{1}{n}\sum\limits_{i=1}^n w(x_i;\theta) s(x_i;\theta)}{\frac{1}{n}\sum\limits_{i=1}^n w(x_i;\theta)} = \frac{\mathbb{E}_{P_{\theta}}[w(X;\theta) s(X;\theta)]}{\mathbb{E}_{P_{\theta}}[w(X;\theta)]},
\end{eqnarray*}
which expands to
\begin{eqnarray}
\label{p2:eqn:density_power_normalized_estimating_equation}
 \frac{\displaystyle \sum\limits_{i=1}^n P_{\theta}(x_i)^c s(x_i;\theta)}{\displaystyle \sum\limits_{i=1}^n P_{\theta}(x_i)^c} = \frac{\displaystyle \sum\limits_{x \in \X} P_{\theta}(x)^{c+1} s(x;\theta)}{\displaystyle \sum\limits_{x \in \X} P_{\theta}(x)^{c+1}}.
\end{eqnarray}
Jones et al. \cite{2001xxBio_Jon_etal} compare the robustness properties of estimators arising from (\ref{p2:eqn:nonnormalized_estimating_equation}) and (\ref{p2:eqn:density_power_normalized_estimating_equation}). According to Jones et al. \cite[p.~866]{2001xxBio_Jon_etal}, the former is more efficient, but the latter has better robustness with respect to a mixture model of contamination with outliers.

Equation (\ref{p2:eqn:density_power_normalized_estimating_equation}) can be recognized as an estimating equation arising from the first order optimality criterion for the maximization
\begin{eqnarray}
\label{p2:mean_power_likelihood}
 \max_{\theta \in \Theta} \left[ \frac{1}{c}\log\left( \frac{1}{n}\sum\limits_{i=1}^n P_{\theta}(x_i)^{c}\right) - \frac{1}{1+c}\log\sum\limits_{x \in \X} P_{\theta}(x)^{1+c} \right].
\end{eqnarray}
We shall soon see why it ought to be a maximization. The objective function in (\ref{p2:mean_power_likelihood}) is called \emph{mean power likelihood} \footnote{To see why the objective function in (\ref{p2:mean_power_likelihood}) is called {\em mean power likelihood}, verify that (\ref{p2:eqn:density_power_normalized_estimating_equation}) is equivalent to
\[
  \frac{1}{n} \sum\limits_{i=1}^n s_c(x_i; \theta) = 0
\]
where
\[
  s_c(x;\theta) := P_{\theta}(x)^c \Big[ s(x;\theta) - \frac{1}{1+c} \nabla_{\theta} \Big(\log\sum\limits_{x \in \X} P_{\theta}(x)^{c+1}\Big) \Big].
\]
The quantity $s_c(x_i; \theta)$ is a generalization of the power-weighted and centered score function. The centering ensures Fisher consistency. As $c \downarrow 0$, we have $s_c(x;\theta) \ra s(x;\theta)$.}. The corresponding estimator is called the maximum mean power likelihood estimate (MMPLE) by Eguchi and Kato \cite{201002Ent_EguKat}; we shall denote it $\hatt_{c+1}$. (The appearance of 1 in the subscript $\hatt_{c+1}$ will soon become clear.) When $c = 0$, we see that $\hatt_1$ becomes the MLE $\hatt$. The parameter $c$ in (\ref{p2:mean_power_likelihood}) can thus be used to trade-off robustness for asymptotic efficiency as observed in \cite{1995xxJRSS_Win}, \cite{2001xxBio_Jon_etal}.

Let us now bring in the connection to a parametric family of relative entropies. Recall that $\hat{P}$ is the empirical measure of the data. The argument $\theta\in\Theta$ that maximizes the objective in (\ref{p2:mean_power_likelihood}) is the same as minimizing
\begin{eqnarray}
  \nonumber
  - \frac{c+1}{c}\log\left(\frac{1}{n}\sum\limits_{i=1}^n P_{\theta}(x_i)^{c}\right) + \frac{1}{c}\log\sum\limits_{x \in \X} \hat{P}(x)^{c+1} + \log\sum\limits_{x \in \X} P_{\theta}(x)^{c+1} \\
  \nonumber
  & & \hspace{-8cm} = - \frac{c+1}{c}\log \sum\limits_{x \in \X} \hat{P}(x)P_{\theta}(x)^c  + \frac{1}{c}\log\sum\limits_{x \in \X} \hat{P}(x)^{c+1} + \log\sum\limits_{x \in \X} P_{\theta}(x)^{c+1} \\
  \label{p2:eqn:first-Ialpha}
  & & \hspace{-8cm} =: \mathscr{I}_{c + 1}(\hat{P},P_{\theta}),
\end{eqnarray}
where $\mathscr{I}_{c+1}$ in (\ref{p2:eqn:first-Ialpha}) is a parametric extension of relative entropies already studied in our companion paper \cite{2014xxManuscript1_KumSun}. We thus have
\begin{equation}
  \label{p2:eqn:I-alpha-projection}
  \hatt_{c+1} = \arg \min_{\theta \in \Theta} \mathscr{I}_{c+1}(\hat{P}, P_{\theta}),
\end{equation}
and the probability measure $P_{\hatt_{c+1}}$ corresponding to the MMPLE $\hatt_{c+1}$ is called the {\em reverse $\mathscr{I}_{c+1}$-projection of the empirical measure $\hat{P}$ on the family $\mathbb{E}$}. It is known (see for example \cite[Lemma~1-b)]{2014xxManuscript1_KumSun}) that $\lim_{c \downarrow 0} \mathscr{I}_{c+1}(P,Q) = \mathscr{I}(P\|Q)$, as it should be, for we already saw that $c = 0$ yields $\hatt_1 = \hatt$, the MLE, which is also the reverse $\I$-projection of the empirical measure $\hat{P}$ on $\mathbb{E}$. This operational continuity intuitively suggests that we must have minimization in (\ref{p2:eqn:I-alpha-projection}) and maximization in (\ref{p2:mean_power_likelihood}).

Let us now use large sample asymptotics to justify the minimization in (\ref{p2:eqn:I-alpha-projection}) (and maximization in (\ref{p2:mean_power_likelihood})). Let $\theta^*$ be the true parameter and let $x_1, \ldots, x_n$ be drawn independently and according to $P_{\theta^*}$. As the number of samples $n$ goes to infinity, almost surely, the empirical measure\footnote{The dependence of $\hat{P}$ on $n$ is understood and suppressed.} $\hat{P}$ converges (point-wise) to the true probability measure $P_{\theta^*}$. For a fixed candidate estimate $\theta$, by virtue of the continuity of $\mathscr{I}_{c+1}(\cdot, P_{\theta})$ in the first argument  when $c > 0$, see \cite[Prop.~2]{2014xxManuscript1_KumSun}, we have (almost surely)
\[
  \I_{c+1}(\hat{P}, P_{\theta}) \stackrel{n \ra \infty}{\rightarrow} \I_{c+1}(P_{\theta^*}, P_{\theta}) \geq \I_{c+1}(P_{\theta^*}, P_{\theta^*}),
\]
where the last inequality follows from the fact that $\I_{\alpha}(P_{\theta^*}, P_{\theta}) \geq 0$ with equality if and only if $\theta = \theta^*$ \cite[Lem.~1-a)]{2014xxManuscript1_KumSun}. From this, it is clear that one must minimize over $\theta \in \Theta$ (and not maximize) in (\ref{p2:eqn:I-alpha-projection}) in order to identify the true parameter $\theta^*$.

Some historical remarks are now called for. Basu et al. \cite{1998xxBIO_Bas} studied a nonnormalized version of the estimating equation (\ref{p2:eqn:density_power_normalized_estimating_equation}), namely (\ref{p2:eqn:nonnormalized_estimating_equation}) with $w(x;\theta) = P_{\theta}(x)^c$. They also identified an associated divergence which is now called $\beta$-divergence \cite{2008xxJma_FujEgu}, \cite{2010xxEnt_CicAma}. The $\beta$-divergences belong to the class of Bregman divergences \cite{1991xxTAS_Csi}. Jones et al. \cite{2001xxBio_Jon_etal} proposed the normalized estimating equation (\ref{p2:eqn:density_power_normalized_estimating_equation}) and identified a divergence associated with (\ref{p2:eqn:density_power_normalized_estimating_equation}), see \cite[Eq.~(2.8)]{2001xxBio_Jon_etal}. Fujisawa and Eguchi \cite{2008xxJma_FujEgu} found that $\mathscr{I}_{c+1}$ is another divergence associated with the estimating equation (\ref{p2:eqn:density_power_normalized_estimating_equation}) and termed it $\gamma$-divergence. They also established an approximate Pythagorean relation for $\mathscr{I}_{c+1}$ (which is quite different from what we shall discuss in Section \ref{p2:sec:forwardprojection}) and used it to bound the error between estimates arising with and without contamination by outliers\footnote{The outliers are generated using a mixture model.}. Recently, Cichocki and Amari \cite{2010xxEnt_CicAma} surveyed the properties of the $\beta$- and the $\Ia$-divergences and their connection to other divergences.

Earlier Sundaresan \cite{200206ISIT_Sun} and \cite{200701TIT_Sun} arrived at $\Ia$-divergences in the context of redundancy in compression and guessing problems (for $\alpha<1$). Let us now turn to this.

\subsection{Reverse $\Ia$-projection: $\alpha < 1$}
\label{p2:subsec:reverse-I-alpha-l1}

We now motivate reverse $\Ia$-projection for $\alpha < 1$. R\'enyi entropies play a role similar to Shannon entropy when one wishes to minimize the normalized cumulant of compressed lengths as opposed to expected compressed lengths. More precisely, with $\rho = \alpha^{-1}-1 >0$, Campbell \cite{1965xxIC_Cam} showed that
\[
  \min \frac{1}{n\rho} \log \mathbb{E}[\exp \{ \rho L_n(X^n) \}] \ra H_{\alpha}(\hat{P}) ~ (\mbox{as $n \ra \infty$})
\]
for an i.i.d. source with marginal $\hat{P}$. The minimization is taken over all length functions $L_n$ that satisfy the Kraft inequality. $\rho$ is the cumulant parameter. As $\alpha\uparrow 1$, we have $\rho\downarrow 0$, and it is well known that $\lim_{\alpha \uparrow 1} H_{\alpha}(\hat{P}) = H(\hat{P})$, the Shannon entropy, so that R\'enyi entropy can be viewed as an operational generalization of Shannon entropy.

Suppose now that the compressor is forced to use for compression, not the true probability measure $\hat{P}$, but a probability measure $P_{\theta}$ from a family parameterized by $\theta \in \Theta$. Let us denote, as before, $\mathbb{E} = \{ P_{\theta} \colon \theta \in \Theta \}$. As an example, $\hat{P}$ may be a generic measure on $\X = \{ 0, 1, \ldots, L \}$, but the compressor may wish to pick the best representation of $\hat{P}$ among binomial distributions $P_{\theta}$ having $\theta \in (0,1)$ as parameter\footnote{More sophisticated examples are possible. Take $\X = \{0,1\}^{\mathbb{Z}}$, $\hat{P}$ any fixed, stationary, and ergodic probability measure on $\X$, and $\mathbb{E}$ the class of stationary Markov measures on $\X$ of fixed Markov order. Since this $\X$ is not finite, such examples are beyond the scope of this paper.}. If the compressor picks $P_{\theta}$ instead of the true $\hat{P}$, then the gap in the resulting normalized cumulant from the optimal value is $\Ia(\hat{P},P_{\theta})$ \cite{200701TIT_Sun}. It follows that the best compressor from within $\mathbb{E}$ has parameter
\begin{equation}
  \label{p2:eqn:I-alpha-projection-1}
  \hatt_{\alpha} = \arg \min_{\theta \in \Theta} \Ia(\hat{P}, P_{\theta})
\end{equation}
and the probability measure $P_{\hatt_{\alpha}}$ is the reverse $\Ia$-projection of $\hat{P}$ on the family $\mathbb{E}$. While (\ref{p2:eqn:I-alpha-projection}) defines reverse $\Ia$-projection for $\alpha > 1$, (\ref{p2:eqn:I-alpha-projection-1}) defines such a projection for $\alpha < 1$. As one expects, $\lim_{\alpha \uparrow 1} \Ia(\hat{P},P_{\theta}) = \I(\hat{P}\|P_{\theta})$, the penalty for mismatch in compression when {\em expected lengths} are considered, and one has the operational continuity that $\I(\hat{P}\|P_{\theta})$ is the usual limiting penalty for mismatch as $\alpha \uparrow 1$.

$\Ia$ also arises as the gap from optimality due to mismatch in performance of guessing schemes (Arikan \cite{199601TIT_Ari}, Hanawal and Sundaresan \cite{201101TIT_HanSun}, Sundaresan \cite{200701TIT_Sun}) and more recently in the performance of coding for tasks (Bunte and Lapidoth \cite{2014xxarx_BunLap}).

\section{The Setting and Contributions}
\label{p2:sec:setting}

In this section, we formalize the notions of projections and the families of interest. We then highlight our contributions.

We begin by recalling the definition of $\mathscr{I}_{\alpha}$ and its alternate expressions.

\vspace*{.1in}

\begin{definition}
The {\em relative $\alpha$-entropy} of $P$ with respect to $Q$ is defined as
\begin{eqnarray}
\mathscr{I}_{\alpha}(P,Q)\label{p2:alphadiv_expanded}
& := & \frac{\alpha}{1-\alpha} \log \Big[ \sum\limits_x P(x) Q(x)^{\alpha-1} \Big] - \frac{1}{1-\alpha}\log \sum\limits_x P(x)^{\alpha} + \log \sum\limits_x Q(x)^{\alpha}\\ \label{p2:alphadiv_linear_form}
& = & \frac{\alpha}{1-\alpha} \log \left[ \sum\limits_x \frac{P(x)}{\|P\|} \left( \frac{Q(x)}{\|Q\|} \right)^{\alpha-1} \right],
\end{eqnarray}
where
\[
 \|Q\| = \Big[ \sum\limits_x Q(x)^{\alpha} \Big]^{1/\alpha}.
\]
\end{definition}

\vspace*{.1in}

Equation (\ref{p2:alphadiv_expanded}) is the same as (\ref{p2:eqn:first-Ialpha}) but with the parameter space extended to $\alpha > 0, \alpha \neq 1$. Equation (\ref{p2:alphadiv_linear_form}) follows after regrouping of terms using the definition of $\|P\|$ and $\|Q\|$. For any $\tau > 0$, since $Q/\|Q\| = \tau Q / \| \tau Q\|$, it follows that (\ref{p2:alphadiv_linear_form}) can be extended to any pair of positive measures $P$ and $Q$ on $\X$, and not just probability measures on $\X$. 

For each $\alpha >0, \alpha\neq 1$, $\Ia(P,Q)\ge 0$ with equality iff $P=Q$.

Note that $\mathscr{I}_{\alpha}(P,Q) = \infty$ if and only if either
\begin{itemize}
  \item $\alpha < 1$ and $P$ is not absolutely continuous with respect to $Q$ (notation $P \not\ll Q$), or
  \item $\alpha > 1$ and $P$ and $Q$ are singular, i.e., the supports of $P$ and $Q$ are disjoint.
\end{itemize}

Let $\mathcal{P}(\mathbb{X})$ be the set of all probability measures on $\mathbb{X}$. For a probability measure $P$ on $\mathbb{X}$, let $\text{Supp}(P)=\{x:P(x)>0\}$ denote the support of $P$. For a set $\mathbb{E}$ of probability measures, write $\text{Supp}(\mathbb{E})$ for the union of the supports of the members of $\mathbb{E}$

Let us now formally define what we mean by a reverse $\Ia$-projection for $\alpha > 0$, $\alpha \neq 1$.

\vspace*{.1in}

\begin{definition}[Reverse $\mathscr{I}_{\alpha}$-projection]
  \label{p2:defn:rev-proj}
  Let $R$ be a probability measure on $\X$. Let $\mathbb{E}$ be a set of probability measures on $\X$ such that $\Ia(R,P) < \infty$ for some $P \in \mathbb{E}$. A probability measure $Q \in \mathbb{E}$ satisfying
  \begin{equation}
    \label{p2:eqn:I-alpha-rev-projection}
    \Ia(R,Q) = \inf_{P \in \mathbb{E}} \Ia(R,P) =: \Ia(R,\mathbb{E})
  \end{equation}
  is called a {\em reverse $\Ia$-projection} of $R$ on $\mathbb{E}$. If there is no such $Q \in \mathbb{E}$, a probability measure $Q$ in the closure of $\mathbb{E}$ satisfying (\ref{p2:eqn:I-alpha-rev-projection}) is called a {\em generalized reverse $\Ia$-projection} of $R$ on $\mathbb{E}$.
\end{definition}
\vspace*{.1in}

In a previous paper \cite{2014xxManuscript1_KumSun}, we studied the {\em forward $\Ia$-projection} of a probability measure $R$ on a family. We reproduce \cite[Defn.~6]{2014xxManuscript1_KumSun} here for it plays a crucial role in this paper.

\vspace*{.1in}
\begin{definition}[Forward $\mathscr{I}_{\alpha}$-projection]
  \label{p2:defn:fwd-proj}
  Let $R$ be a probability measure on $\X$. Let $\mathbb{E}$ be a set of probability measures on $\X$ such that $\Ia(P,R) < \infty$ for some $P \in \mathbb{E}$. A probability measure $Q \in \mathbb{E}$ satisfying
  \begin{equation}
    \label{p2:eqn:I-alpha-fwd-projection}
    \Ia(Q,R) = \inf_{P \in \mathbb{E}} \Ia(P,R) =: \Ia(\mathbb{E},R)
  \end{equation}
  is called a {\em forward $\Ia$-projection} of $R$ on $\mathbb{E}$.
\end{definition}
\vspace*{.1in}

In Definition \ref{p2:defn:rev-proj}, the minimization is with respect to the second argument, while in Definition \ref{p2:defn:fwd-proj} the minimization is with respect to the first argument. The focus in \cite{2014xxManuscript1_KumSun} was on forward projection on convex families and general alphabet spaces. We provided sufficient conditions for existence of the forward projection and argued that if the forward projection exists then it is unique. Convex families arise naturally from constraints placed by measurements of linear statistics. Examples of such families are linear families which we now define.

\vspace*{.1in}
\begin{definition}[Linear family]
\label{p2:defn:linear-family}
A linear family characterized by $k$ functions $f_i: \X \ra \R$, $1 \leq i \leq k$, is the set of probability measures given by
\begin{eqnarray}
\label{p2:eqn:linear_family}
 \mathbb{L} := \Big\{P\in \mathcal{P}(\mathbb{X})\colon \sum\limits_x P(x)f_i(x) = 0, i = 1,\dots,k\Big\}.
\end{eqnarray}
\end{definition}
\vspace*{.1in}

Reverse $\Ia$-projections, however, correspond to maximum likelihood or robust estimations, and are often on exponential families which we now define.

\vspace*{.1in}
\begin{definition}[Exponential family]
\label{p2:defn:expo-family}
An exponential family characterized by a probability measure $R$ and $k$ functions $f_i\colon \X \ra \R$, $1 \leq i \leq k$, is the set of probability measures given by
\[
  \mathbb{M} := \left\{ P_{\theta} \colon  \theta \in \Theta \subset \R^k \right\},
\]
where
\begin{eqnarray*}
  P_{\theta}(x)^{-1} & := & Z(\theta) \exp \Big[ \log \left( R(x)^{-1} \right) + \sum\limits_{i=1}^k \theta_i f_i(x) \Big] \\
                & = & Z(\theta) R(x)^{-1} \exp \Big[ \sum\limits_{i=1}^k \theta_i f_i(x) \Big] \quad \forall x \in \X
\end{eqnarray*}
with $Z(\theta)$ being the normalization constant and $\Theta$ being the subset of $\R^k$ for which $P_{\theta}$ is a valid probability measure\footnote{If $R(x)$ equals $0$, then so does $P_{\theta}(x)$.}.
\end{definition}
\vspace*{.1in}

Examples of exponential families include
\begin{itemize}
 \item Bernoulli distribution ($\X = \{0,1\}$, $\Theta = (0,1)$),
 \item Binomial distribution ($\X = \{0,1,\ldots, L\}$, $\Theta=(0,1))$,
 \item Poisson distribution ($\X = \{0,1,\ldots\}$, $\Theta = (0,\infty)$), and
 \item Gaussian distribution $(\X = \R^d$, the parameter $\theta$ denotes the pair of mean and covariance).
\end{itemize}
The last two are given only as illustrative examples for they do not satisfy the finite $\X$ assumption of this paper. We will take up the study of reverse $\Ia$-projection on the more general {\em log-convex} families which we now define.

\vspace*{.1in}
\begin{definition}[Log-convex family]
\label{p2:defn:log-convex}
A set $\mathbb{E}$ of probability measures on a finite alphabet set $\X$ is said to be \emph{log-convex} if for any two probability measures $P$ and $Q$ in $\mathbb{E}$ that are not singular, and any $t\in [0,1]$, the probability measure $\overline{P^t Q^{1-t}}$ defined by
\begin{equation}
  \label{p2:eqn:log-convex-mixture}
  \overline{P^t Q^{1-t}}(x) := \frac{P(x)^t Q(x)^{1-t}}{\sum\limits_y P(y)^t Q(y)^{1-t}}
\end{equation}
also belongs to $\mathbb{E}$.
\end{definition}
\vspace*{.1in}

Exponential families are log-convex, a fact that is easily checked.

We will also take up reverse projections on analogs of exponential families. To define these analogs, let us first define the generalized logarithm and the generalized exponential functions \cite{199406QN_Tsa}. Let $\bar{\mathbb{R}}_{+} = \mathbb{R}\cup \{+\infty\}$ and let $\bar{\mathbb{R}} = \mathbb{R}\cup \{+\infty, -\infty\}$.

\vspace*{.1in}
\begin{definition}
\label{p2:defn:gen-log-exp}
For $\alpha > 0$, the $\alpha$-logarithm function, denoted $\ln_{\alpha}\colon\bar{\mathbb{R}}_{+} \ra \bar{\mathbb{R}}$, is defined to be
\[
  \ln_{\alpha} (u) := \left\{ \begin{array}{ll}
    \frac{u^{1-\alpha} - 1}{1-\alpha} & \alpha \neq 1 \\
    \log (u) & \alpha = 1
  \end{array}
  \right.
\]
where the log function is the natural logarithm. Its functional inverse, the $\alpha$-exponential function, denoted $e_{\alpha}\colon\bar{\mathbb{R}} \ra \bar{\mathbb{R}}_{+}$, is defined to be
\[
  e_{\alpha} (u) := \left\{ \begin{array}{ll}
    (\max\{ 1+(1-\alpha)u ,0\})^{1/(1-\alpha)} & \alpha \neq 1 \\
    \exp (u) & \alpha = 1.
  \end{array}
  \right.
\]
\end{definition}
\vspace*{.1in}

It is easy to check that $e_{\alpha}(\ln_{\alpha}(u)) = u$ for $u > 0$ and that $\ln_{\alpha}(e_{\alpha}(u)) = u$ whenever $0 < e_{\alpha}(u) < \infty$.

The analogs of exponential families are the so-called $\alpha$-power-law families which we now define. (Compare Definitions \ref{p2:defn:expo-family} and \ref{p2:defn:alpha-power-family}.)

\vspace*{.1in}
\begin{definition}[$\alpha$-power-law family]
\label{p2:defn:alpha-power-family}
Let $R$ be a probability measure such that if $\alpha >1$ then $\text{Supp}(R) = \mathbb{X}$. An $\alpha$-power-law family characterized by the probability measure $R$ and $k$ functions $f_i\colon \X \ra \R$, $1 \leq i \leq k$, is the set of probability measures given by
\[
  \mathbb{M}^{(\alpha)} := \left\{ P_{\theta} \colon  \theta \in \Theta \subset \R^k \right\},
\]
where
\begin{eqnarray}
\label{p2:eqn:power-law1}
  P_{\theta}(x)^{-1} := Z(\theta) e_{\alpha} \Big[ \ln_{\alpha} \left( R(x)^{-1} \right) + \sum\limits_{i=1}^k \theta_i f_i(x) \Big] \quad \forall x\in \mathbb{X},
\end{eqnarray}
provided
\[
 1 + (1-\alpha) \Big[\ln_{\alpha} \left( R(x)^{-1} \right) + \sum\limits_{i=1}^k \theta_i f_i(x)\Big] >0 \quad \forall x \in \X,
\]
with $Z(\theta)$ being the normalization constant and $\Theta$ being the subset of $\R^k$ for which $P_{\theta}$ is a valid probability measure.
\end{definition}
Equivalently\footnote{A definition such as (\ref{p2:eqn:power-law1}) is fraught with pesky issues of well-definedness. We have verified the equivalence of (\ref{p2:eqn:power_law_family}). But a skeptical reader may simply take (\ref{p2:eqn:power_law_family}) as the starting point to define $\mathbb{M}^{(\alpha)}$. The definition in (\ref{p2:eqn:power-law1}) is given only to highlight its similarity with Definition \ref{p2:defn:expo-family}. Observe that, from (\ref{p2:eqn:power_law_family}), if $\alpha  <1$, $R(x)=0$ implies $P_{\theta}(x)=0$.}, 
\begin{eqnarray}
\label{p2:eqn:power_law_family}
  P_{\theta}(x)^{\alpha -1} = Z(\theta)^{1-\alpha} \Big[R(x)^{\alpha-1} + (1-\alpha)\sum\limits_{i=1}^k \theta_i f_i(x)\Big] >0 \quad \forall x\in \mathbb{X}.
\end{eqnarray}
When we wish to be explicit about the characterizing entities, we shall write $\mathbb{M}^{(\alpha)}(R, f_1, \dots, f_k)$ for the family.
In Appendix \ref{p2:app:weakdependence}, we show that $\mathbb{M}^{(\alpha)}$ depends on $R$ in only a weak manner. Any member $P_{\theta^*} \in \mathbb{M}^{(\alpha)}$ may equally well play the role of $R$ and this merely corresponds to translation and scaling of the parameter space.

$\mathbb{M}^{(\alpha)}$ is not closed. Sometimes it will be required to consider its closure $\text{cl}(\mathbb{M}^{(\alpha)})$.

One has the more general notion of $\ln_{\alpha}$-convex family as well (see van Erven and Harremo\"es \cite{201407TIT_ErvHar}\footnote{van Erven and Harremo\"es \cite{201407TIT_ErvHar} gave a different name to what we call $\ln_{\alpha}$-convex family; they called this $(\alpha-1)$-convex family. Our convention follows the notation for and parametrization of the generalized logarithm.}).

\vspace*{.1in}
\begin{definition}[$\ln_{\alpha}$-convex family]
\label{p2:defn:ln-alpha-convex}
A set $\mathbb{E}$ of probability measures is said to be \emph{$\ln_{\alpha}$-convex} if for any two probability measures $P$ and $Q$ in $\mathbb{E}$ (that are not singular when $\alpha \leq 1$), and any $t\in [0,1]$, the probability measure $R$ defined by
\begin{equation}
  \label{p2:eqn:ln-alpha-combination}
  R^{-1} := Z e_{\alpha}\left( t \ln_{\alpha}(P^{-1}) + (1-t) \ln_{\alpha}(Q^{-1}) \right)
\end{equation}
also belongs to $\mathbb{E}$. The quantity $Z$ is the normalization constant that makes $R$ a probability measure.
\end{definition}

\vspace*{.1in}

Substitution of the definitions of $e_{\alpha}$ and $\ln_{\alpha}$ indicate that the probability measure $R$ defined in (\ref{p2:eqn:ln-alpha-combination}) can be rewritten as
\begin{equation}
  \label{p2:eqn:ln-alpha-combination-1}
  Z^{-1} \left[ t P^{\alpha-1} + (1-t) Q^{\alpha-1} \right]^{\frac{1}{\alpha-1}}.
\end{equation}
When $\alpha=1$, $\ln_{\alpha}$-convexity is just log-convexity, thereby justifying that $\ln_{\alpha}$-convexity is an extension of log-convexity. Just as exponential families are log-convex, $\alpha$-power-law families are $\ln_{\alpha}$-convex, a fact that can be easily checked using (\ref{p2:eqn:ln-alpha-combination-1}).

While forward projections of interest are on convex families, reverse projections of interest, particularly those arising in estimation problems, are on log-convex, and by analogy, on $\ln_{\alpha}$-convex families. Log-convex or $\ln_{\alpha}$-convex families are not necessarily convex in the usual sense.

\vspace*{.1in}

Definition \ref{p2:defn:ln-alpha-convex} is given only to complete the picture. We shall restrict attention in this paper to the $\alpha$-power-law family.
\subsection{A closer look at our contributions.}
\label{p2:subsec:contributions}

For a given $R$ and a given $\mathbb{E}$ with some $P$ such that $\Ia(R,P) < \infty$, we obviously have $\Ia(R,\mathbb{E}) < \infty$. If we consider a sequence $(P_n) \subset \mathbb{E}$ such that $\lim_{n \ra \infty} \Ia(R,P_n) = \Ia(R,\mathbb{E})$, by virtue of the continuity of $\Ia(P, \cdot)$ in the second argument (see \cite[Rem.~5]{2014xxManuscript1_KumSun}), all subsequential limits of $(P_n)$ are generalized reverse $\Ia$-projections. In this paper, we study example settings when the generalized reverse $\Ia$-projection is unique, when it is not, and how one may characterize it, sometimes, as a forward $\Ia$-projection. Specifically, we do the following.

\begin{itemize}
  \item In Section \ref{p2:sec:reverse_projection}, we study reverse $\Ia$-projections on log-convex families. We show an example of nonuniqueness of generalized reverse $\Ia$-projections on an exponential family when $\alpha > 1$. However uniqueness holds for $\alpha < 1$.

  \item In Section \ref{p2:sec:forwardprojection}, our focus will be on the forward $\mathscr{I}_{\alpha}$-projection on certain convex families, in particular, linear families. We identify the form of the forward $\mathscr{I}_{\alpha}$-projection on a linear family $\mathbb{L}$ and prove a necessary and sufficient condition for a $Q\in\mathbb{L}$ to be the forward $\mathscr{I}_{\alpha}$-projection on $\mathbb{L}$. We consider the cases $\alpha > 1$ and $\alpha < 1$ separately in two subsections. The proof for the $\alpha <1$ case is similar to Csisz\'{a}r and Shields' proof for $\alpha =1$ case \cite{2004xxITST_CsiShi}. For the proof of the $\alpha >1$ case, we resort to the Lagrange multiplier technique. The structure of the forward $\mathscr{I}_{\alpha}$-projection naturally suggests a statistical model, namely the $\alpha$-power-law family $\mathbb{M}^{(\alpha)}$.

  \item In Section \ref{p2:sec:orthogonality}, we study reverse $\Ia$-projections on $\mathbb{M}^{(\alpha)}$, and show uniqueness of the generalized reverse projection for all $\alpha > 0, \alpha \neq 1$. To show this, we establish an orthogonality relationship between $\mathbb{M}^{(\alpha)}$ and an associated linear family. We then use this geometric property to turn a reverse $\Ia$-projection on $\mathbb{M}^{(\alpha)}$ into a forward $\Ia$-projection on the linear family. It will turn out that, sometimes, we may need to consider a larger family than just $\text{cl}(\mathbb{M}^{(\alpha)})$.
\end{itemize}

\section{Reverse projection onto log-convex sets}
\label{p2:sec:reverse_projection}

We consider the cases $\alpha > 1$ and $\alpha < 1$ separately in the next two subsections. Before that, we present a lemma of some independent interest. This is an extension of a result for relative entropy ($\alpha = 1$); see Csisz\'ar and Mat\'{u}\v{s} \cite[Eq.~(3)]{200306TIT_CsiMat}, where (\ref{p2:eqn:log-convex}) below is an equality.

\vspace*{.1in}
\begin{lemma}
\label{p2:lemma:logconvex}
Let $P$ and $Q$ be probability measures on $\mathbb{X}$ that are mutually absolutely continuous. Let $R$ be any probability measure on $\mathbb{X}$ that is not singular with respect to $P$ or $Q$. Let $t \in [0,1]$.
\begin{itemize}
  \item[(a)] If $\alpha < 1$, then
  \begin{eqnarray}
     \label{p2:eqn:log-convex}
     t \mathscr{I}_{\alpha}(R,P)+(1-t) \mathscr{I}_{\alpha}(R,Q) \ge \mathscr{I}_{\alpha}(R,\overline{P^t Q^{1-t}}) - \log \sum\limits_x P'(x)^t Q'(x)^{1-t},
  \end{eqnarray}
  where $P'$ is the escort probability measure associated with $P$ given by
  \[
     P'(x) := \frac{P(x)^{\alpha}}{\sum\limits_y P(y)^{\alpha}}
  \]
  and $Q'$ is the escort probability measure associated with $Q$.

  \item[(b)] If $\alpha > 1$, the inequality in (\ref{p2:eqn:log-convex}) is reversed.
\end{itemize}
\end{lemma}
\vspace*{.1in}

\begin{IEEEproof}
Let us first observe that if $\alpha < 1$ and $R \not\ll \overline{P^t Q^{1-t}}$, then, by the assumption that $P$ and $Q$ are mutually absolutely continuous, both sides of (\ref{p2:eqn:log-convex}) are $+\infty$, and so (\ref{p2:eqn:log-convex}) holds. We may thus assume that $R \ll \overline{P^t Q^{1-t}}$ when $\alpha < 1$. Also, notice that the hypotheses imply that $R$ is not singular with respect to $\overline{P^t Q^{1-t}}$. Hence, for both $\alpha <1$ and $\alpha >1$, we may take all the terms in (\ref{p2:eqn:log-convex}) to be finite.

Let us write
\begin{eqnarray*}
\frac{P(x)^t Q(x)^{1-t}}{\sum\limits_y P(y)^t Q(y)^{1-t}} =
\frac{\left(\frac{P(x)}{\|P\|}\right)^t \left(\frac{Q(x)}{\|Q\|}\right)^{1-t}}{\sum\limits_y \left(\frac{P(y)}{\|P\|}\right)^t \left(\frac{Q(y)}{\|Q\|}\right)^{1-t}}.
\end{eqnarray*}
Using this in (\ref{p2:alphadiv_linear_form}) we get
\begin{eqnarray*}
\mathscr{I}_{\alpha}(R,\overline{P^t Q^{1-t}})
 & = & \frac{\alpha}{1-\alpha}\log \sum\limits_x \frac{R(x)}{\|R\|} \left(\frac{\Big(\frac{P(x)}{\|P\|}\Big)^t \Big(\frac{Q(x)}{\|Q\|}\Big)^{1-t}}{\left(\sum\limits_y \Big(\frac{P(y)}{\|P\|}\Big)^{\alpha t} \Big(\frac{Q(y)}{\|Q\|}\Big)^{\alpha (1-t)}\right)^{\frac{1}{\alpha}}}\right)^{\alpha-1}\\
& = & \frac{\alpha}{1-\alpha}\log \sum\limits_x \frac{R(x)}{\|R\|} \left[\left(\frac{P(x)}{\|P\|}\right)^t \left(\frac{Q(x)}{\|Q\|}\right)^{1-t}\right]^{\alpha-1} + \log \sum\limits_x \left(\frac{P(x)}{\|P\|}\right)^{\alpha t} \left(\frac{Q(x)}{\|Q\|}\right)^{\alpha (1-t)}\\
& = & \frac{\alpha}{1-\alpha}\log \sum\limits_x \left[\frac{R(x)}{\|R\|} \left(\frac{P(x)}{\|P\|}\right)^{\alpha-1}\right]^t \left[\frac{R(x)}{\|R\|}\left(\frac{Q(x)}{\|Q\|}\right)^{\alpha-1}\right]^{1-t} + \log \sum\limits_x P'(x)^t Q'(x)^{1-t}\\
& \le & \frac{\alpha}{1-\alpha}\log \left[\sum\limits_x \frac{R(x)}{\|R\|} \left(\frac{P(x)}{\|P\|}\right)^{\alpha-1}\right]^t \left[\sum\limits_x \frac{R(x)}{\|R\|}\left(\frac{Q(x)}{\|Q\|}\right)^{\alpha-1}\right]^{1-t} + \log \sum\limits_x P'(x)^t Q'(x)^{1-t}\\
& = & t \mathscr{I}_{\alpha}(R,P)+(1-t) \mathscr{I}_{\alpha}(R,Q) + \log \sum\limits_x P'(x)^t Q'(x)^{1-t},
\end{eqnarray*}
for $\alpha <1$, where the penultimate inequality follows by applying H\"{o}lder's inequality to the inner-product within the first logarithm term, with exponents $1/t$ and $1/(1-t)$. For $\alpha >1$, the inequality is obviously reversed because the multiplication factor $\alpha/(1-\alpha)$ is negative.
\end{IEEEproof}

\subsection{Reverse $\Ia$-projection for $\alpha > 1$}

Recall that the MMPLE on a log-convex family is the reverse $\Ia$-projection of the empirical measure on the family for the case when $\alpha > 1$. For log-convex families, it is possible that multiple reverse $\Ia$-projections may exist, and we provide an explicit example.

\vspace*{.1in}
\begin{example}
\label{p2:ex:multiple-reverse-projections}
  Let $\X = \{0,1,2\}$, let $R$ be the uniform probability measure on $\X$, and let $\mathbb{E}$ be the log-convex family of binomial distributions on $\X$ with parameter $\theta \in (0,1)$. A member $P_{\theta}$ of the family is given by
  \[
    P_{\theta}(0) = (1-\theta)^2, ~P_{\theta}(1) = 2 \theta(1-\theta), ~P_{\theta}(2) = \theta^2.
  \]


  \begin{center}
  \begin{figure*}[t]
    \centering
    \label{p2:fig:multipleprojections}
    \includegraphics[scale=0.5]{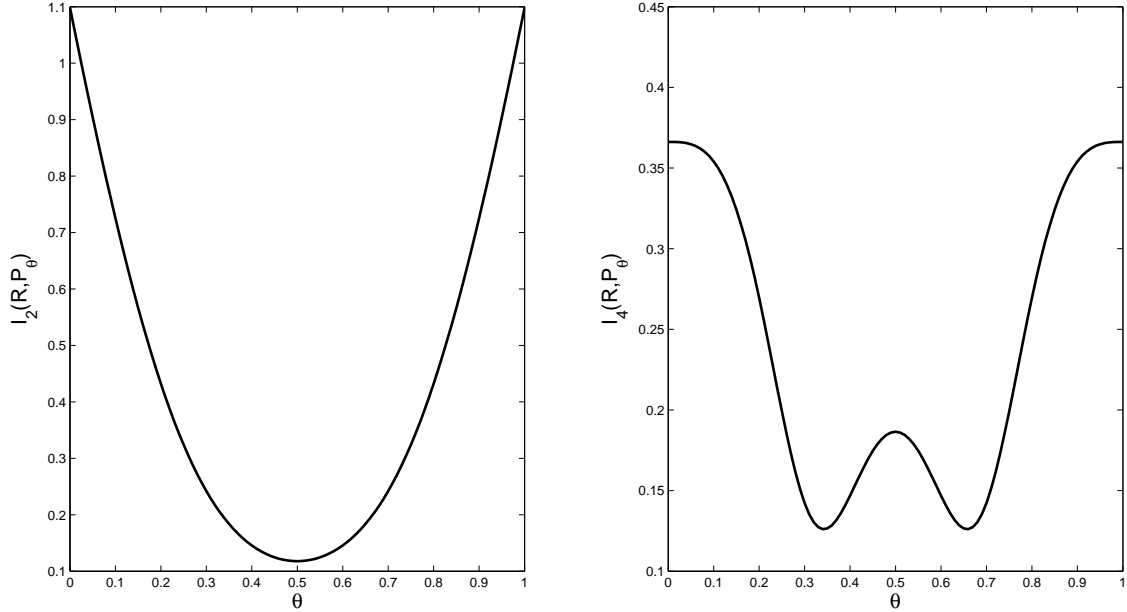}
    \caption{Multiple reverse $\Ia$-projections are possible when $\alpha > 1$.}
  \end{figure*}
  \end{center}
  Figure \ref{p2:fig:multipleprojections} plots $\Ia(R,P_{\theta})$ as a function of $\theta$ for $\alpha = 2$ (plot on the left-hand side) and $\alpha = 4$ (plot on the right-hand side). Since $\Ia(R,P_{\theta})$ has mirror-symmetry around the point $\theta = 1/2$, a fact that can be easily checked, if there is a global minimum at $\theta^* \in (0,\frac{1}{2})$, then we have another global minimum at $1-\theta^* \in (\frac{1}{2},1)$. This is the situation with the plot on the right-hand side.
\end{example}
\vspace*{.1in}

Eguchi and Kato \cite{201002Ent_EguKat} consider the problem of {\em spontaneous} clustering for a Gaussian mixture model with an unknown number of components, and put the possibility of multiple minima to good use. Very briefly, their procedure operates on the data as follows, and we refer the interested reader to \cite{201002Ent_EguKat} for further details. They first choose the parameter $\alpha$ with some care using either the maximum range of the data or the Akaike information criterion. They then identify the resulting minima of $\Ia(R,P_{\theta})$ over the parameters $\theta \in \Theta$. Here $R$ is the empirical measure\footnote{The empirical measure $R$ and the Gaussian $P_{\theta}$ are singular. Following the formal definition in \cite[Sec.~II]{2014xxManuscript1_KumSun}, strictly speaking, we have the relative $\alpha$-entropy $\Ia(R,P_{\theta}) = \infty$. The expansion however does provide a valid expression for optimization although one cannot interpret it as the relative $\alpha$-entropy, and Eguchi and Kato \cite{201002Ent_EguKat} minimize the expression to get the MMPLE.} of the data and $\alpha$ is as chosen. They interpret each minimum point as the parameter of a ``discovered'' component of the mixture. Finally, they associate each data point to a nearby component, among those discovered, thereby arriving at a clustering. If the number of components is unknown, the number of minima is a {\em spontaneous} choice for the number of components of the mixture.

Example \ref{p2:ex:multiple-reverse-projections} suggests a sequence $(P_n) \subset \mathbb{E}$ that satisfies $\Ia(R, P_n) \ra \Ia(R,\mathbb{E})$, and yet $P_n$ does not converge: take $\alpha = 4$, $P_n = P_{\theta^*}$ for odd $n$, and $P_n = P_{1-\theta^*}$ for even $n$. All subsequential limits are of course generalized reverse $\Ia$-projections.

\subsection{Reverse $\Ia$-projection for $\alpha < 1$}

For $\alpha < 1$, the generalized reverse $\Ia$-projection is unique, unlike the situation in the previous subsection.

\vspace{0.1in}

\begin{theorem}
\label{p2:thm:gen_reverse}
Let $\alpha <1$. Let $\mathbb{E}$ be a log-convex set of mutually absolutely continuous probability measures on $\mathbb{X}$. Let $R$ be a probability measure on $\mathbb{X}$ such that $\mathscr{I}_{\alpha}(R,\mathbb{E}) < \infty$. Under these conditions, there exists a unique probability measure $Q$ such that, for every sequence $(P_n)$ in $\mathbb{E}$ satisfying $\mathscr{I}_{\alpha}(R,P_n)\to \mathscr{I}_{\alpha}(R,\mathbb{E})$,  we have $P_n\to Q$ and $\mathscr{I}_{\alpha}(R,Q) = \mathscr{I}_{\alpha}(R,\mathbb{E})$.
\end{theorem}
\vspace*{.1in}

\begin{IEEEproof}
The proof broadly follows the proof of Csisz\'ar's \cite[Th. 1]{200306TIT_CsiMat}.

Consider a sequence $(P_n) \subset \mathbb{E}$ such that $\lim_n \mathscr{I}_{\alpha}(R,P_n) = \mathscr{I}_{\alpha}(R,\mathbb{E})$. Since $\mathscr{I}_{\alpha}(R,\mathbb{E})$ is finite, we may assume without loss of generality that $\mathscr{I}_{\alpha}(R,P_n)$ is finite for all $n$. Hence, for all $n$, $R$ is not singular with respect to $P_n$; indeed, $R \ll P_n$ for all $n$. Apply Lemma \ref{p2:lemma:logconvex} with $P=P_m$, $Q=P_n$ to get
\begin{eqnarray}
t \mathscr{I}_{\alpha}(R,P_m) + (1-t) \mathscr{I}_{\alpha}(R,P_n) & \ge & \mathscr{I}_{\alpha}(R,\overline{P_m^t P_n^{1-t}}) - \log \sum\limits_x P_m'(x)^t P_n'(x)^{1-t}\\
\label{p2:eqn:gen-rev-proj-inequality}
& \ge & \mathscr{I}_{\alpha}(R,\mathbb{E}) - \log \sum\limits_x P_m'(x)^t P_n'(x)^{1-t},
\end{eqnarray}
where last inequality follows from the hypothesis that $\overline{P_m^t P_n^{1-t}} \in \mathbb{E}$. Also observe that, by H\"older's inequality,
\begin{equation}
  \label{p2:eqn:Holder-consequence}
  \sum\limits_x P_m'(x)^t P_n'(x)^{1-t} \leq \Big(\sum\limits_x P_m'(x)\Big)^t \Big(\sum\limits_x P_n'(x)\Big)^{1-t} = 1.
\end{equation}
Let $m,n\to \infty$ in (\ref{p2:eqn:gen-rev-proj-inequality}) and use (\ref{p2:eqn:Holder-consequence}) to get
\[
 \lim_{m,n \to \infty} \log \sum\limits_x P_m'(x)^t P_n'(x)^{1-t} = 0.
\]
Set $t = 1/2$ in this limit and undo the logarithm to get
\[
  \lim_{m,n \to \infty} \sum\limits_x \sqrt{P_m'(x) P_n'(x)} = 1
\]
so that
\begin{eqnarray*}
\sum\limits_x\left(\sqrt{P_m'(x)} - \sqrt{P_n'(x)}\right)^2 & = & 2-2\cdot \sum\limits_x \sqrt{P_m'(x) P_n'(x)}\\
& \to & 0 \, \, \, \, \mbox{ as } m,n \to \infty.
\end{eqnarray*}
Thus $(P_n')$ is a Cauchy sequence. It must converge to some $Q'$, an escort of some probability measure $Q$. Given our finite alphabet assumption, we must then have $P_n \to Q$.

If $(Q_n) \subset \mathbb{E}$ is another sequence such that $\mathscr{I}_{\alpha}(R,Q_n)\to \mathscr{I}_{\alpha}(R,\mathbb{E})$, then since $P_n$ and $Q_n$ can be merged together, $(Q_n)$ must also converge to the same $Q$. The generalized reverse $\Ia$-projection is therefore unique.

By continuity of $\mathscr{I}_{\alpha}(R,\cdot)$, see \cite[Rem. 5]{2014xxManuscript1_KumSun}, we also have $\mathscr{I}_{\alpha}(R,Q) = \mathscr{I}_{\alpha}(R,\mathbb{E})$.
\end{IEEEproof}
\vspace{0.1in}

The proof fails for $\alpha > 1$ because the inequality in (\ref{p2:eqn:gen-rev-proj-inequality}) is in the opposite direction, and one cannot conclude that $(P_n')$ is a Cauchy sequence. Indeed, the previous subsection provides a counterexample for lack of convergence and nonuniqueness of reverse $\Ia$-projection on a log-convex family, when $\alpha > 1$.

\vspace{0.1in}

\section{Forward $\mathscr{I}_{\alpha}$-projection}
\label{p2:sec:forwardprojection}
In this section, we will recall some results on forward $\mathscr{I}_{\alpha}$-projection from \cite{2014xxManuscript1_KumSun} along with some refinements for our restricted finite alphabet setting. The proofs here use elementary tools and exploit the finite alphabet assumption. The results will then be used to turn a reverse $\mathscr{I}_{\alpha}$-projection on an $\alpha$-power-law family into a forward $\mathscr{I}_{\alpha}$-projection on a linear family.

 \subsection{$\alpha <1$:}

The result for $\alpha <1$ is the following. It establishes the form of the forward $\mathscr{I}_{\alpha}$-projection on a linear family.

\begin{figure}[tb]
\centering
 \includegraphics[height = 3cm, width = 4cm]{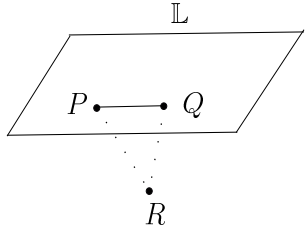}
\caption{\label{fig:pythagorean}Pythagorean property}
\end{figure}

\vspace{0.1in}
\begin{theorem}
 \label{p2:thm:forwardprojection_alpha<1}
Let $\alpha <1$. Let $\mathbb{L}$ be a linear family characterized by $f_i, i=1, \dots, k$. Let $R$ be a probability measure with full support. Then the following hold.
\begin{itemize}
 \item[(a)] $R$ has a forward $\mathscr{I}_{\alpha}$-projection on $\mathbb{L}$. Call it $Q$.

 \item[(b)] $\text{Supp}(Q) = \text{Supp}(\mathbb{L})$ and the Pythagorean equality holds (see Figure \ref{fig:pythagorean}):
\begin{eqnarray}
 \label{p2:eqn:pythagorean_equality}
  \mathscr{I}_{\alpha}(P,R) = \mathscr{I}_{\alpha}(P,Q) + \mathscr{I}_{\alpha}(Q,R) \quad \forall P \in \mathbb{L}.
\end{eqnarray}

 \item[(c)] The forward $\mathscr{I}_{\alpha}$-projection $Q$ satisfies 
\begin{eqnarray}
 \label{p2:eqn:power-law2}
Z^{\alpha -1} Q(x)^{\alpha - 1} = R(x)^{\alpha -1} + (1-\alpha)\sum\limits_{i=1}^k \theta_i^* f_i(x) \quad \forall x \in \text{Supp}(\mathbb{L}),
\end{eqnarray}
where $\theta_1^*, \dots, \theta_k^*$ are scalars and $Z$ is the normalization constant that makes $Q$ a probability measure.

 \item[(d)] The forward $\mathscr{I}_{\alpha}$-projection is unique.
\end{itemize}
\end{theorem}

\vspace{0.1in}

\begin{IEEEproof}
(a) The mapping $P \mapsto \mathscr{I}_{\alpha}(P,R)$ is continuous \cite[Rem. 5]{2014xxManuscript1_KumSun} and $\mathbb{L}$ is compact. Hence the forward $\mathscr{I}_{\alpha}$-projection exists.

(b) This follows from \cite[Props. 14-15, Th. 10-a]{2014xxManuscript1_KumSun}.

(c) Our proof follows the proof of Csisz\'{a}r and Shields proof for the case $\alpha =1$ \cite[Th. 3.2]{2004xxITST_CsiShi}.

From (\ref{p2:eqn:linear_family}), it is clear that the probability measures $P\in \mathbb{L}$, when considered as $|\text{Supp}(\mathbb{L})|$-dimensional vectors, belong to the orthogonal complement $\mathcal{F}^{\perp}$ of the subspace $\mathcal{F}$ of $\mathbb{R}^{|\text{Supp}(\mathbb{L})|}$ spanned by the vectors $f_i(\cdot), i = 1, \dots, k$, restricted to $\text{Supp}(\mathbb{L})$. These $P\in \mathbb{L}$ actually span $\mathcal{F}^{\perp}$. (This follows from the fact that if a subspace of $\mathbb{R}^{|\text{Supp}(\mathbb{L})|}$ contains a vector all of whose components are strictly positive, here $Q$, then it is spanned by the probability vectors of that space.) Using (\ref{p2:alphadiv_linear_form}), one can see (\ref{p2:eqn:pythagorean_equality}) same as
\[
 \sum\limits_x P(x)\Bigg(\frac{R(x)^{\alpha-1}}{\sum\limits_a Q(a)R(a)^{\alpha-1}} - \frac{Q(x)^{\alpha-1}}{\sum\limits_a Q(a)^{\alpha}}\Bigg) = 0 \quad \forall P\in \mathbb{L}.
\]
Consequently, the vector
\[
 \frac{R(\cdot)^{\alpha-1}}{\sum\limits_a Q(a)R(a)^{\alpha-1}} - \frac{Q(\cdot)^{\alpha-1}}{\sum\limits_a Q(a)^{\alpha}}
\]
belongs to $(\mathcal{F}^{\perp})^{\perp} = \mathcal{F}$, that is,
\[
 \frac{R(x)^{\alpha-1}}{\sum\limits_a Q(a)R(a)^{\alpha-1}} - \frac{Q(x)^{\alpha-1}}{\sum\limits_a Q(a)^{\alpha}} = \sum_{i=1}^k \lambda_i f_i(x) \quad \forall x\in \text{Supp}(\mathbb{L})
\]
for some scalars $\lambda_i, i=1, \dots, k$ . This verifies (\ref{p2:eqn:power-law2}) for obvious choices of $Z$ and $\theta_i^*$.

(d) This follows from \cite[Th. 8]{2014xxManuscript1_KumSun}.
\end{IEEEproof}

\vspace{0.1in}

One can also state a converse.

\vspace{0.1in}

\begin{theorem}
\label{p2:thm:forwardprojection_alpha<1_converse}
 Let $\alpha <1$. Let $Q \in \mathbb{L}$ be a probability measure of the form (\ref{p2:eqn:power-law2}). Then 
$Q$ satisfies (\ref{p2:eqn:pythagorean_equality}) and is the forward $\mathscr{I}_{\alpha}$-projection of $R$ on $\mathbb{L}$.
\end{theorem}

\vspace{0.1in}

\begin{IEEEproof}
 This follows from \cite[Th. 11-b]{2014xxManuscript1_KumSun}.
\end{IEEEproof}
 
 \subsection{$\alpha >1$:}
   \label{p2:sec:forwardprojection_alpha>1}
\vspace{0.1in}
We now establish the form of the forward $\mathscr{I}_{\alpha}$-projection on a linear family when $\alpha >1$. The following result may be seen as a refinement of \cite[Th. 10(a)]{2014xxManuscript1_KumSun}.
\vspace{0.1in}

\begin{theorem}
 \label{p2:thm:forwardprojection_alpha>1}
Let $\alpha >1$. Let $\mathbb{L}$ be a linear family characterized by $f_i,i=1, \dots, k$. Let $R$ be a probability measure with full support. Then the following hold.
\begin{itemize}
 \item[(a)] $R$ has a forward $\mathscr{I}_{\alpha}$-projection on $\mathbb{L}$. Call it $Q$.

 \item[(b)] The forward $\mathscr{I}_{\alpha}$-projection $Q$ satisfies 
\begin{equation}
 \label{p2:eqn:power_law_with+}
Z^{\alpha -1} Q(x)^{\alpha - 1}  =  \Big[R(x)^{\alpha -1} + (1-\alpha)\sum\limits_{i=1}^k \theta_i^* f_i(x)\Big]_{+} \quad \forall x \in \mathbb{X},
\end{equation}
where $\theta_1^*, \dots, \theta_k^*$ are scalars, $Z$ is the normalization constant that makes $Q$ a probability measure, and $[u]_{+} = \max\{u,0\}$.

\item[(c)] The Pythagorean inequality holds:
\begin{eqnarray}
 \label{p2:eqn:pythagorean_inequality1}
  \mathscr{I}_{\alpha}(P,R) \ge \mathscr{I}_{\alpha}(P,Q) + \mathscr{I}_{\alpha}(Q,R) \quad \forall P \in \mathbb{L}.
\end{eqnarray}

\item[(d)] The forward $\mathscr{I}_{\alpha}$-projection is unique.

\item[(e)] If $\text{Supp}(Q) = \text{Supp}(\mathbb{L})$, then (\ref{p2:eqn:pythagorean_inequality1}) holds with equality.
\end{itemize}

\end{theorem}
\vspace{0.1in}

\begin{IEEEproof}
(a) The mapping $P \mapsto \mathscr{I}_{\alpha}(P, R)$ is continuous \cite[Prop. 2]{2014xxManuscript1_KumSun} and $\mathbb{L}$ is compact. Hence the forward $\mathscr{I}_{\alpha}$-projection exists.

(b) The optimization problem for the forward $\mathscr{I}_{\alpha}$-projection is
\begin{align}
 \min_P \, & \mathscr{I}_{\alpha}(P,R)\label{p2:min}\\
\mbox{subject to } &\sum\limits_x P(x)f_i(x) = 0, \quad i=1, \dots, k \label{p2:eqn:linear_constraints}\\
                   &\sum\limits_x P(x)       = 1 \label{p2:eqn:probability_constraint}\\
                   &P(x)                    \ge 0 \quad \forall x \in \mathbb{X}. \label{p2:eqn:positivity_constraints}
\end{align}
We will proceed in a sequence of steps.
\begin{itemize}
 \item[(i)] Observe that $\mathscr{I}_{\alpha}(\cdot,R)$, in addition to being continuous, is also continuously differentiable. Indeed, we have
\begin{equation}
\label{p2:eqn:partial_derivative}
 \frac{\partial}{\partial P(x)}\mathscr{I}_{\alpha}(P,R) = \frac{\alpha}{1-\alpha}\Bigg[\frac{R(x)^{\alpha - 1}}{\sum\limits_a P(a)R(a)^{{\alpha-1}}} - \frac{P(x)^{\alpha - 1}}{\sum\limits_a P(a)^{\alpha}}\Bigg].
\end{equation}
Both denominators are bounded away from zero because for any $P \in \mathbb{L}$, we have $\max_x P(x) \ge 1/|\mathbb{X}|$, and therefore 
\[
 \sum\limits_a P(a)R(a)^{\alpha-1} \ge \frac{1}{|\mathbb{X}|}\cdot \min_a R(a)^{\alpha-1} >0,
\]
and
\[
 \sum\limits_a P(a)^{\alpha} \ge \frac{1}{|\mathbb{X}|^{\alpha}} >0.
\]
Consequently, the partial derivative (\ref{p2:eqn:partial_derivative}) exists everywhere on $\mathbb{R}_{+}^{|\mathbb{X}|}$, and is continuous because the terms involved are continuous. (The numerator of the second term in (\ref{p2:eqn:partial_derivative}) is continuous because $\alpha >1$).

 \item[(ii)] Since the equality constraints in (\ref{p2:eqn:linear_constraints}) and (\ref{p2:eqn:probability_constraint}) arise from affine functions, and the inequality constraints in (\ref{p2:eqn:positivity_constraints}) arise from linear functions, we may apply \cite[Prop. 3.3.7]{2003xxNLP_Ber} to conclude that there exist Lagrange multipliers ($\lambda_i, i=1, \dots, k$), $\nu$, and $\left(\mu(x),x\in \mathbb{X}\right)$ associated with the constraints (\ref{p2:eqn:linear_constraints}), (\ref{p2:eqn:probability_constraint}), and (\ref{p2:eqn:positivity_constraints}), respectively, that satisfy:
\begin{align}
\label{p2:eqn:lagrange1}
 \frac{\alpha}{1-\alpha} \Bigg[\frac{Q(x)^{\alpha - 1}}{\sum\limits_a Q(a)^{\alpha}} - \frac{R(x)^{\alpha - 1}}{\sum\limits_a Q(a)R(a)^{\alpha-1}}\Bigg] & = \sum\limits_{i=1}^k \lambda_i f_i(x) - \mu(x) + \nu \quad \forall x\\
 \label{p2:eqn:feasibility_condition}
\mu(x) & \ge 0 \quad \forall x\\
\label{p2:eqn:slackness_condition}
\mu(x)Q(x) & = 0 \quad \forall x.
\end{align}
In writing (\ref{p2:eqn:lagrange1}), we have substituted (\ref{p2:eqn:partial_derivative}) for $\frac{\partial}{\partial P(x)}\mathscr{I}_{\alpha}(P,R)$.

\item[(iii)] Multiplying (\ref{p2:eqn:lagrange1}) by $Q(x)$, summing over all $x\in \mathbb{X}$, using $Q\in \mathbb{L}$, and using (\ref{p2:eqn:slackness_condition}), we see that $\nu = 0$.

\item[(iv)] If $Q(x) >0$, we must have $\mu(x) =0$ from (\ref{p2:eqn:slackness_condition}), and its substitution in (\ref{p2:eqn:lagrange1}) yields, for all such $x$,
\begin{eqnarray}
\label{p2:eqn:lagrange2}
 \frac{Q(x)^{\alpha - 1}}{\sum\limits_a Q(a)^{\alpha}} = \frac{R(x)^{\alpha - 1}}{\sum\limits_a Q(a)R(a)^{\alpha-1}} + \frac{1-\alpha}{\alpha}\sum\limits_{i=1}^k \lambda_i f_i(x).
\end{eqnarray}
If $Q(x) =0$, (\ref{p2:eqn:lagrange1}) implies that
\begin{equation}
\label{p2:eqn:lagrange3}
\frac{R(x)^{\alpha - 1}}{\sum\limits_a Q(a)R(a)^{\alpha-1}} + \frac{1-\alpha}{\alpha}\sum\limits_{i=1}^k \lambda_i f_i(x)
 = \frac{(1-\alpha)}{\alpha} \mu(x) \le 0,
\end{equation}
where the last inequality holds because of (\ref{p2:eqn:feasibility_condition}) and $\alpha >1$.
Therefore, (\ref{p2:eqn:lagrange2}) and (\ref{p2:eqn:lagrange3}) may be combined as 
\begin{eqnarray*}
Z^{\alpha -1} Q(x)^{\alpha - 1} = \Big[R(x)^{\alpha -1} + (1-\alpha)\sum\limits_{i=1}^k \theta_i^* f_i(x)\Big]_{+} \quad \forall x \in \mathbb{X},
\end{eqnarray*}
where the choices of $Z$ and $\theta_i^*$ are obvious. This verifies (\ref{p2:eqn:power_law_with+}) and completes the proof of (b).
\end{itemize}

(c) This follows from \cite[Th. 10-a]{2014xxManuscript1_KumSun}.

(d) Follows from \cite[Th. 8]{2014xxManuscript1_KumSun}.

(e) This can be shown using the proof of \cite[Prop. 15]{2014xxManuscript1_KumSun} and using \cite[Th. 10-a]{2014xxManuscript1_KumSun}.
\end{IEEEproof}

\vspace{0.1in}

As in the $\alpha <1$ case, one has a converse.

\vspace{0.1in}

\begin{theorem}
\label{p2:thm:forwardprojection_alpha>1_converse}
 Let $\alpha >1$. Let $Q \in \mathbb{L}$ be a probability measure of the form (\ref{p2:eqn:power_law_with+}). Then 
$Q$ satisfies (\ref{p2:eqn:pythagorean_inequality1}) for every $P\in \mathbb{L}$, and $Q$ is the forward $\mathscr{I}_{\alpha}$-projection of $R$ on $\mathbb{L}$.
\end{theorem}

\vspace{0.1in}

\begin{IEEEproof}
Follows from \cite[Th. 11-b]{2014xxManuscript1_KumSun}.
\end{IEEEproof}

\vspace{0.1in}

When $\alpha >1$, in general, $\text{Supp}(Q) \neq \text{Supp}(\mathbb{L})$ as shown by the following counterexample, and the Pythagorean inequality (\ref{p2:eqn:pythagorean_inequality1}) may be strict.

\vspace{0.1in}

\begin{example}
\label{p2:eg:not_alg_inner_point}
 Let $\alpha = 2$. Let $\X = \{ 1,2,3,4 \}$. Write $P = (p_1, p_2, p_3, p_4)$ for a probability measure on $\X$. Define the linear family $\mathbb{L}$ to be
\begin{equation*}
  \mathbb{L} = \left\{ P\in \mathcal{P}(\X) \colon 8 p_1 + 4 p_2 + 2 p_3 + p_4 = 7\right\}.
\end{equation*}
Let $R$ be the uniform probability measure on $\X$. We claim that the forward $\mathscr{I}_{\alpha}$-projection of $R$ on $\mathbb{L}$ is $Q = (\nicefrac{3}{4}, \nicefrac{1}{4}, 0, 0)$.
\end{example}
First, $Q\in \mathbb{L}$ because $8 q_1 + 4 q_2 + 2 q_3 + q_4 = 8\times \nicefrac{3}{4} + 4\times \nicefrac{1}{4} + 0 + 0 = 7$. Second, $Q$ is of the form (\ref{p2:eqn:power_law_with+}). To see this, let us note that $f_1(\cdot) = (1,-3,-5,-6)$. Take $\theta_1^* = -\nicefrac{1}{20}$ and $Z=\nicefrac{2}{5}$. Then 
\begin{eqnarray*}
\left[R(\cdot)^{\alpha-1}+(1-\alpha)\theta_1^* f_1(\cdot)\right]_{+}
& = & [R(\cdot) - \theta_1^* f_1(\cdot)]_{+}\\
& = & ([\nicefrac{1}{4}+\nicefrac{1}{20}]_{+},[\nicefrac{1}{4}-\nicefrac{3}{20}]_{+},[\nicefrac{1}{4}-\nicefrac{5}{20}]_{+}, [\nicefrac{1}{4}-\nicefrac{6}{20}]_{+})\\
& = & (\nicefrac{6}{20}, \nicefrac{2}{20},0,0)\\
& = & Z \cdot Q(\cdot).
\end{eqnarray*}
That $Q$ is the forward $\mathscr{I}_{\alpha}$-projection now follows from Theorem \ref{p2:thm:forwardprojection_alpha>1_converse}.

Clearly $\text{Supp}(Q) \subsetneqq \text{Supp}(\mathbb{L})$. Also for $P = (0.8227,0.0625,0.0536,0.0612)\in\mathbb{L}$, numerical calculations yield a strict inequality in (\ref{p2:eqn:pythagorean_inequality1}) since the left-hand side and the right-hand side of (\ref{p2:eqn:pythagorean_inequality1}) evaluate to $1.0114$ and $0.9871$, respectively. See also \cite[Rem.~13]{2014xxManuscript1_KumSun} where this counterexample showed that transitivity of projections does not hold for $\alpha > 1$. In both situations, the issue is that $\text{Supp}(Q) \neq \text{Supp}(\mathbb{L})$.

%

\section{Orthogonality between the $\alpha$-power-law family and the linear family}
\label{p2:sec:orthogonality}
\begin{figure}[tb]
\centering
 \includegraphics[height = 4cm, width = 4cm]{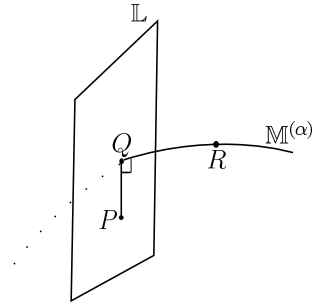}
\caption{\label{p2:fig:orthogonal_intersection} Orthogonal intersection of an $\alpha$-power-law family and a linear family}
\end{figure}

The focus of this section is on the geometry of the $\alpha$-power-law family with respect to its associated linear family, and its exploitation. See Figure \ref{p2:fig:orthogonal_intersection}. We treat the cases $\alpha <1$ and $\alpha >1$ separately. Theorems \ref{p2:thm:compression_problem} and \ref{p2:thm:mmple} are the main contributions.

 \subsection{$\alpha <1$:}
  \label{p2;sec:orthogonality_alpha<1}

\vspace{0.1in}

  This case is the simpler of the two. The core result of this section, one on which the main result Theorem \ref{p2:thm:compression_problem} hinges, is the following that shows that the case $\alpha <1$ is similar to $\alpha =1$ \cite[Th. 3.2]{2004xxITST_CsiShi}.

\vspace{0.1in}

\begin{thm}
\label{p2:thm:orthogonal_intersection_alpha<1}
Let $\alpha < 1$. Let $\mathbb{L}$ be a linear family characterized by $f_i, i=1, \dots, k$, as in (\ref{p2:eqn:linear_family}). Let $R$ be a probability measure with full support. Let $\mathbb{M}^{(\alpha)}$ be the $\alpha$-power-law family, as in Definition \ref{p2:defn:alpha-power-family}, characterized by $R$ and the same $k$ functions $f_i, i=1, \dots, k$. Let $Q$ be the forward $\mathscr{I}_{\alpha}$-projection of $R$ on $\mathbb{L}$. Then the following hold.
\begin{itemize}
 \item[(a)] $\mathbb{L}\cap \text{cl}(\mathbb{M}^{(\alpha)}) = \{Q\}$.

 \item[(b)] For every $P \in \mathbb{L}$, we have
\begin{eqnarray}
\label{p2:eqn:pythagorean_equality1}
 \mathscr{I}_{\alpha}(P, R) = \mathscr{I}_{\alpha}(P, Q) + \mathscr{I}_{\alpha}(Q, R).
\end{eqnarray}
 \item[(c)] If $\text{Supp}(\mathbb{L}) = \mathbb{X}$, then $\mathbb{L}\cap \mathbb{M}^{(\alpha)} = \{Q\}$.
\end{itemize}
\end{thm}

\vspace{0.1in}

\begin{IEEEproof}
Statement (b) is the same as Theorem \ref{p2:thm:forwardprojection_alpha<1}-(c). Let us observe from Theorem \ref{p2:thm:forwardprojection_alpha<1} that when $\text{Supp}(\mathbb{L}) = \mathbb{X}$, the forward $\Ia$-projection $Q$ of $R$ on $\mathbb{L}$ satisfies
\[
 Z^{\alpha -1} Q(x)^{\alpha -1} = R(x)^{\alpha-1} + (1-\alpha) \sum\limits_{i=1}^k\theta_i^* f_i(x) \quad \forall x\in \mathbb{X}
\]
for some scalars $Z, \theta_1^*,\dots,\theta_k^*$. Hence $Q \in \mathbb{M}^{(\alpha)}$. Since $Q$ is also in $\mathbb{L}$, we have $Q \in \mathbb{L}\cap \mathbb{M}^{(\alpha)}$.

Thus, in general, when $\text{Supp}(\mathbb{L})$ is not necessarily $\mathbb{X}$, if we can show that (i) every member of $\mathbb{L}\cap \text{cl}(\mathbb{M}^{(\alpha)})$ satisfies (\ref{p2:eqn:pythagorean_equality1}), and (ii) $\mathbb{L}\cap \text{cl}(\mathbb{M}^{(\alpha)})$ is nonempty, then, since any member satisfying (\ref{p2:eqn:pythagorean_equality1}) is also forward $\mathscr{I}_{\alpha}$-projection and since the forward $\Ia$-projection is unique, the theorem will be established. We now proceed to show (i) and (ii).
\begin{eqnarray}
  \label{p2:intersection_satisfies_pyth}
  \hspace{-10.5cm}\mbox{(i) Every } \tilde Q\in \mathbb{L}\cap \text{cl}(\mathbb{M}^{(\alpha)}) \mbox{ satisfies }  (\ref{p2:eqn:pythagorean_equality1}).
\end{eqnarray}
Let $(Q_n) \subset \mathbb{M}^{(\alpha)}$ be such that $Q_n \to \tilde Q$. Then, for each $n$, there exist $\theta^{(n)} = \left(\theta_1^{(n)},\dots,\theta_k^{(n)}\right) \in \mathbb{R}^k$ and a constant $Z_n$ such that
\begin{equation}
\label{p2:power-law-of-Q_n}
 Z_n^{\alpha-1} Q_n(x)^{\alpha-1} = R(x)^{\alpha-1} + (1-\alpha) \sum\limits_{i=1}^k\theta_i^{(n)} f_i(x) \quad \forall x\in \mathbb{X}.
\end{equation}
Since, for any $P\in \mathbb{L}$, we have
\[
\sum\limits_x P(x) f_i(x) = \sum\limits_x \tilde Q(x)f_i(x) = 0, \quad i=1, \dots, k,
\]
by taking expectation with respect to $P$ and $\tilde Q$ on both sides of (\ref{p2:power-law-of-Q_n}), we get
\[
 Z_n^{\alpha-1} \sum\limits_x P(x) Q_n(x)^{\alpha-1} =  \sum\limits_{x} P(x) R(x)^{\alpha-1}
\]
and
\[
 Z_n^{\alpha-1} \sum\limits_x \tilde Q(x) Q_n(x)^{\alpha-1} = \sum\limits_{x} \tilde Q(x) R(x)^{\alpha-1},
\]
respectively. Using the above two equations to eliminate $Z_n^{\alpha-1}$, we get
\[
 \sum\limits_x P(x)R(x)^{\alpha-1} = \frac{\sum\limits_x \tilde Q(x) R(x)^{\alpha-1}}{\sum\limits_x \tilde Q(x) Q_n(x)^{\alpha-1}}\sum\limits_x P(x) Q_n(x)^{\alpha-1}.
\]
Letting $n\to \infty$, and then by using (\ref{p2:alphadiv_expanded}), we get (\ref{p2:eqn:pythagorean_equality1}) with $Q$ replaced by $\tilde Q$. This proves (i).

\vspace{0.1in}

(ii) $\mathbb{L}\cap \text{cl}(\mathbb{M}^{(\alpha)})$ is nonempty.

\vspace{0.1in}

Let
\begin{align*}
 \tau_i^{(n)} & := \frac{\frac{1}{n} \sum\limits_x R(x)f_i(x)}{(1-\frac{1}{n})\sum\limits_x Q(x)R(x)^{\alpha-1} + \frac{1}{n}\sum\limits_x R(x)^{\alpha}},\\
\tilde f_i(\cdot) & := f_i(\cdot) - \tau_i^{(n)} R(\cdot)^{\alpha-1},  \quad i = 1, \dots, k,
\end{align*}
and define the sequence of linear families
\[
 \mathbb{L}_n := \Big\{P\in \mathcal{P}(\mathbb{X}) \colon \sum\limits_x P(x) \tilde f_i(x) = 0, \, i = 1,\dots, k \Big\}.
\]
The $\tau_i^{(n)}$'s are chosen so that $(1-\frac{1}{n})Q + \frac{1}{n}R \in \mathbb{L}_n$, and so $\text{Supp}(\mathbb{L}_n) = \mathbb{X}$. Let $Q_n$ be the forward $\mathscr{I}_{\alpha}$-projection of $R$ on $\mathbb{L}_n$. Then, by virtue of Theorem \ref{p2:thm:forwardprojection_alpha<1}-(b), we have $\text{Supp}(Q_n) = \mathbb{X}$, and by virtue of Theorem \ref{p2:thm:forwardprojection_alpha<1}-(c), we have
\begin{eqnarray}
\label{p2:eqn:ftildetof}
Z_n^{\alpha-1} Q_n(x)^{\alpha-1} & = & R(x)^{\alpha-1} + (1-\alpha) \sum\limits_{i=1}^k\theta_i^{(n)} \tilde f_i(x)\nonumber\\
 & = & \Big[1- (1-\alpha) \sum\limits_{i=1}^k \theta_i^{(n)} \tau_i^{(n)}\Big] R(x)^{\alpha-1} + (1-\alpha) \sum\limits_{i=1}^k\theta_i^{(n)} f_i(x) \quad \forall x \in \mathbb{X}.
\end{eqnarray}
Taking expectation with respect to $Q$ on both sides, and using $\sum\limits_x Q(x)f_i(x) = 0, \, i=1, \dots, k$, we get
\begin{eqnarray*}
Z_n^{\alpha-1} \sum\limits_x Q(x) Q_n(x)^{\alpha-1} = \Big[1- (1-\alpha) \sum\limits_{i=1}^k \theta_i^{(n)} \tau_i^{(n)}\Big] \cdot \sum\limits_x Q(x) R(x)^{\alpha-1}.
\end{eqnarray*}
As the summations on either side are finite and strictly positive for each $n$, the term within square brackets in the above equation is also strictly positive for each $n$. Rescaling (\ref{p2:eqn:ftildetof}) appropriately, we see that $Q_n \in \mathbb{M}^{(\alpha)}$. Note also that $\tau_i^{(n)}\to 0$ as $n\to \infty$ for $i = 1,\dots,k$. Hence the limit of any convergent subsequence of $(Q_n)$ belongs to $\mathbb{L}\cap \text{cl}(\mathbb{M}^{(\alpha)})$. This verifies (ii) and concludes the proof of the theorem.
\end{IEEEproof}

\vspace{0.1in}

We now argue that the family $\text{cl}(\mathbb{M}^{(\alpha)})$ and $\mathbb{L}$ are ``orthogonal'' to each other, in a sense made precise in the statement of the next result.

\vspace{0.1in}

\begin{corollary}
\label{p2:cor:orthogonal_intersection_alpha<1}
Under the hypotheses of Theorem \ref{p2:thm:orthogonal_intersection_alpha<1}, the following additional statements hold.
\begin{itemize}
 \item[(a)] For every $P \in \mathbb{L}$ and every $S \in \text{cl}(\mathbb{M}^{(\alpha)})$, we have
\begin{eqnarray}
\label{p2:eqn:pythagorean_equality2}
 \mathscr{I}_{\alpha}(P, S) = \mathscr{I}_{\alpha}(P, Q) + \mathscr{I}_{\alpha}(Q, S).
\end{eqnarray}
 \item[(b)] For any $S \in \text{cl}(\mathbb{M}^{(\alpha)})$, the forward $\mathscr{I}_{\alpha}$-projection of $S$ on $\mathbb{L}$ is $Q$.
\end{itemize}
\end{corollary}

\vspace{0.1in}

\begin{IEEEproof}
 Since any member of $\mathbb{M}^{(\alpha)}$ can play the role of $R$ by Prop. \ref{p2:app:weakdependence} (in the Appendix), and since, by Theorem \ref{p2:thm:orthogonal_intersection_alpha<1}, $\mathbb{L}\cap \text{cl}(\mathbb{M}^{(\alpha)}) = \{Q\}$, $Q$ is the forward $\mathscr{I}_{\alpha}$-projection of any member of $\mathbb{M}^{(\alpha)}$ on $\mathbb{L}$. Therefore (\ref{p2:eqn:pythagorean_equality2}) holds for every $P \in \mathbb{L}$ and every $S \in \mathbb{M}^{(\alpha)}$. Furthermore, (\ref{p2:eqn:pythagorean_equality2}) holds for the limit of any sequence of members of $\mathbb{M}^{(\alpha)}$, and hence (a) and (b) hold for members of $\text{cl}(\mathbb{M}^{(\alpha)})\setminus \mathbb{M}^{(\alpha)}$ as well.
\end{IEEEproof}

\vspace{0.1in}

Let us now return to the compression problem discussed in Section \ref{p2:subsec:reverse-I-alpha-l1} and show the connection between the reverse $\mathscr{I}_{\alpha}$-projection on an $\alpha$-power-law family and a forward $\mathscr{I}_{\alpha}$-projection on a linear family. 

\vspace{0.1in}

\begin{theorem}
\label{p2:thm:compression_problem}
 Let $\alpha <1$. Let $\hat{P}$ be a probability measure on $\mathbb{X}$. Let $\mathbb{M}^{(\alpha)}$ be characterized by the probability measure $R$ and the functions $f_i, i=1, \dots, k$. Let $\mathbb{L}$ be the associated linear family characterized by $f_i, i=1, \dots, k$, and assume that it is nonempty. Let $R$ have full support.

Define $\tilde{\mathbb{L}}$ as
\begin{equation}
\label{p2:eqn:elltilde}
 \tilde{\mathbb{L}} := \Big\{P\in \mathcal{P}(\mathbb{X}) \colon \sum\limits_{x} P(x) \tilde f_i(x) = 0 \Big\},
\end{equation}
where 
\begin{equation}
\label{p2:eqn:eftilde}
 \tilde f_i(\cdot) = f_i(\cdot) - \tau_i^R R(\cdot)^{\alpha -1}
\end{equation}
with 
\begin{equation}
 \label{p2:eqn:tau}
\tau_i^R = \frac{\sum\limits_x \hat{P}(x)f_i(x)}{\sum\limits_x \hat{P}(x)R(x)^{\alpha -1}}, i = 1, \dots, k.
\end{equation}
Let $Q$ be the forward $\mathscr{I}_{\alpha}$-projection of $R$ on $\tilde{\mathbb{L}}$.
\begin{itemize}
 \item[(a)] If $\text{Supp}(Q) = \mathbb{X}$, then $Q$ is the unique reverse $\mathscr{I}_{\alpha}$-projection of $\hat{P}$ on $\mathbb{M}^{(\alpha)}$.

 \item[(b)] If $\text{Supp}(Q) \neq \mathbb{X}$, then $\hat{P}$ does not have a reverse $\mathscr{I}_{\alpha}$-projection on $\mathbb{M}^{(\alpha)}$. However, $Q$ is the unique reverse $\mathscr{I}_{\alpha}$-projection of $\hat{P}$ on $\text{cl}(\mathbb{M}^{(\alpha)})$.
\end{itemize}
\end{theorem}

\vspace{0.1in}

\begin{IEEEproof}
$\tilde{\mathbb{L}}$ is constructed so that $\hat{P} \in \tilde{\mathbb{L}}$ (which is easy to check) and, further, $\tilde{\mathbb{L}}$ is orthogonal to $\mathbb{M}^{(\alpha)}$ in the sense of Corollary \ref{p2:cor:orthogonal_intersection_alpha<1}. We now verify the latter statement. For concreteness, we will index the the $\alpha$-power-law family by its characterizing entities. By Corollary \ref{p2:cor:orthogonal_intersection_alpha<1}, $\tilde{\mathbb{L}}$ is orthogonal to $\mathbb{M}^{(\alpha)}(R,\tilde f_1, \dots, \tilde f_k)$. It therefore suffices to show that $\mathbb{M}^{(\alpha)}(R,\tilde f_1, \dots, \tilde f_k) = \mathbb{M}^{(\alpha)}(R, f_1, \dots, f_k)$. Take any $P_{\theta} \in \mathbb{M}^{(\alpha)}(R, f_1, \dots, f_k)$. Then, for each $x\in \mathbb{X}$, we have
\begin{eqnarray*}
Z(\theta)^{\alpha -1} P_{\theta}(x)^{\alpha -1} & = & R(x)^{\alpha-1} + (1-\alpha) \sum\limits_{i=1}^k\theta_i f_i(x)\\
& = & (1+(1-\alpha)\theta_i \tau_i^R) R(x)^{\alpha-1} + (1-\alpha) \sum\limits_{i=1}^k\theta_i \tilde f_i(x).
\end{eqnarray*}
Taking expectation with respect to $\hat{P}$ on both sides, and using $\sum_x \hat{P}(x) \tilde f_i(x) = 0, \, i=1, \dots, k$, we get
\[
 Z({\theta})^{\alpha -1} \sum\limits_x \hat{P}(x) P_{\theta}(x)^{\alpha -1} = \left[1+(1-\alpha)\theta_i \tau_i^R\right]\cdot \sum\limits_x \hat{P}(x) R(x)^{\alpha-1}.
\]
Since $P_{\theta}$ and $R$ have full support, it follows that $\left[1+(1-\alpha)\theta_i \tau_i^R\right] > 0$, and hence $P_{\theta} \in \mathbb{M}^{(\alpha)}(R,\tilde f_1, \dots, \tilde f_k)$. This shows $\mathbb{M}^{(\alpha)}(R, f_1, \dots, f_k) \subset \mathbb{M}^{(\alpha)}(R,\tilde f_1, \dots, \tilde f_k)$. Similarly, using the assumption that $\mathbb{L}$ is nonempty, one can show that $\mathbb{M}^{(\alpha)}(R,\tilde f_1, \dots, \tilde f_k) \subset \mathbb{M}^{(\alpha)}(R, f_1, \dots, f_k)$.

By Corollary \ref{p2:cor:orthogonal_intersection_alpha<1}, we have
\begin{equation}
\label{p2:eqn:pythagorean_equality3}
 \mathscr{I}_{\alpha}(\hat{P}, S) = \mathscr{I}_{\alpha}(\hat{P}, Q) + \mathscr{I}_{\alpha}(Q, S) \quad \forall S \in \text{cl}(\mathbb{M}^{(\alpha)}).
\end{equation}
(a) If $\text{Supp}(Q) = \mathbb{X}$, then by Th. \ref{p2:thm:orthogonal_intersection_alpha<1}(c), $Q \in \mathbb{M}^{(\alpha)}$, and from (\ref{p2:eqn:pythagorean_equality3}), the minimum of $\mathscr{I}_{\alpha}(\hat{P}, S)$ over $S\in \mathbb{M}^{(\alpha)}$ is attained at $S = Q$. To prove the uniqueness, let $P_{\theta^*} \in \mathbb{M}^{(\alpha)}$ also attain the minimum. Then, from (\ref{p2:eqn:pythagorean_equality3}), we have
\begin{equation}
\label{p2:eqn:pythagorean_equality4}
 \mathscr{I}_{\alpha}(\hat{P}, P_{\theta^*}) = \mathscr{I}_{\alpha}(\hat{P}, Q) + \mathscr{I}_{\alpha}(Q, P_{\theta^*}).
\end{equation}
Since $\mathscr{I}_{\alpha}(\hat{P}, P_{\theta^*}) = \mathscr{I}_{\alpha}(\hat{P}, Q)$, we have $\mathscr{I}_{\alpha}(Q, P_{\theta^*}) = 0$, and so $P_{\theta^*} = Q$.

(b) Let $\text{Supp}(Q) \neq \mathbb{X}$. Then, by Th. \ref{p2:thm:orthogonal_intersection_alpha<1}(a), $Q\in \text{cl}(\mathbb{M}^{(\alpha)})\setminus \mathbb{M}^{(\alpha)}$. Uniqueness on the closure follows just as in (a) immediately above. If $\hat{P}$ has a reverse $\mathscr{I}_{\alpha}$-projection on $\mathbb{M}^{(\alpha)}$, say $P_{\theta^*}$, then by continuity of $\mathscr{I}_{\alpha}(\hat{P}, \cdot)$ (\cite[Rem. 5]{2014xxManuscript1_KumSun}), we have $\mathscr{I}_{\alpha}(\hat{P}, Q) = \mathscr{I}_{\alpha}(\hat{P}, P_{\theta^*})$. This contradicts the uniqueness.
\end{IEEEproof}

\subsection{$\alpha >1$:}
  \label{p2:sec:orthogonality_alpha>1}

Let us begin with a counterexample that shows that Theorem \ref{p2:thm:orthogonal_intersection_alpha<1} does not hold when $\alpha >1$; $\text{cl}(\mathbb{M}^{(\alpha)})$ need not intersect the associated $\mathbb{L}$.

\vspace{0.1in}

\begin{example} 
Let $\alpha, \mathbb{X}, \mathbb{L}$, and $R$ be as in Example \ref{p2:eg:not_alg_inner_point}. The associated $\alpha$-power-law family and its closure are
\[
 \mathbb{M}^{(\alpha)} = \Big\{P_{\theta} \colon \theta \in (\nicefrac{-1}{24}, \nicefrac{1}{4})\Big\},
\]
and
\[
 \text{cl}(\mathbb{M}^{(\alpha)}) = \Big\{P_{\theta} \colon \theta \in [\nicefrac{-1}{24}, \nicefrac{1}{4}]\Big\},
\]
 where 
\begin{eqnarray*}
 P_{\theta} = \frac{1}{1+13 \theta} \Big( \nicefrac{1}{4}-\theta, \nicefrac{1}{4}+3\theta, \nicefrac{1}{4}+5\theta, \nicefrac{1}{4}+6\theta\Big).
\end{eqnarray*}
We assert that no such $P_{\theta}$, either of $\mathbb{M}^{(\alpha)}$ or $\text{cl}(\mathbb{M}^{(\alpha)})$, is in $\mathbb{L}$. Furthermore, the forward $\mathscr{I}_{\alpha}$-projection of every member in $\text{cl}(\mathbb{M}^{(\alpha)})$ on $\mathbb{L}$ is $Q = (\nicefrac{3}{4}, \nicefrac{1}{4},0,0)$ which, of course, is not in $\text{cl}(\mathbb{M}^{(\alpha)})$. 
\end{example}

\vspace{0.1in}

One must therefore extend $\mathbb{M}^{(\alpha)}$ beyond its closure to identify the family that is orthogonal to $\mathbb{L}$ and intersects $\mathbb{L}$ at $Q$. An appropriate extension of $\mathbb{M}^{(\alpha)}$ that intersects $\mathbb{L}$ turns out to be the following.

\vspace{0.1in}

\begin{definition}
\label{p2:defn:m-alphahat}
 The family $\hat{\mathbb{M}}^{(\alpha)}_{+}$ characterized by a probability measure $R$ and $k$ functions $f_i\colon\mathbb{X}\to \mathbb{R}, i=1, \dots, k$, is defined as follows. Let $Q = P_{\theta^*}$ be the forward $\mathscr{I}_{\alpha}$-projection\footnote{By virtue of Th. \ref{p2:thm:forwardprojection_alpha>1}(b), $Q$ is of the form (\ref{p2:eqn:power_law_with+}) for some $\theta^*$ and hence may be written as $Q = P_{\theta^*}$.} of $R$ on $\mathbb{L}$. Define $\hat{\mathbb{M}}^{(\alpha)}_{+}$ to be the set of all probability measures $P_{\theta}$ satisfying (a), (b), and (c) below.

\begin{itemize}
 \item[(a)] \begin{eqnarray*}
  Z(\theta)^{\alpha-1} P_{\theta}^{\alpha-1}(x) = \Big[R(x)^{\alpha-1} + (1-\alpha)\sum\limits_{i=1}^k \theta_i f_i(x)\Big]_{+} \quad \forall x \in \X,
\end{eqnarray*}
where $Z(\theta)$ is the normalization constant that makes $P_{\theta}$ a valid probability measure on $\mathbb{X}$.

 \item[(b)] $\text{Supp}(P_{\theta^*}) \subseteq \text{Supp}(P_{\theta})$;

 \item[(c)] $\sum\limits_{i=1}^k \theta_i f_i(x) \le \sum\limits_{i=1}^k \theta_i^* f_i(x) \quad \forall x\notin \text{Supp}(P_{\theta})$.
\end{itemize}
\end{definition}

\vspace{0.1in}

The following is the analog of the combined Theorem \ref{p2:thm:orthogonal_intersection_alpha<1} and Corollary \ref{p2:cor:orthogonal_intersection_alpha<1}.
\vspace{0.1in}

\begin{theorem}
\label{p2:thm:orthogonal_intersection_alpha>1}
 Let $\alpha > 1$. Let $\mathbb{L}$ be a linear family characterized by $f_i, i=1, \dots, k$ as in (\ref{p2:eqn:linear_family}). Let $\mathbb{M}^{(\alpha)}$ be as in Definition \ref{p2:defn:alpha-power-family}, characterized by $R$ and the $k$ functions $f_i, i=1, \dots, k$. Let $Q$ be the forward $\mathscr{I}_{\alpha}$-projection of $R$ on $\mathbb{L}$. Let $\hat{\mathbb{M}}^{(\alpha)}_{+}$ be the extension of $\mathbb{M}^{(\alpha)}$ as in Definition \ref{p2:defn:m-alphahat}. We then have the following.
\begin{itemize}
 \item[(a)] $\mathbb{L} \cap \hat{\mathbb{M}}^{(\alpha)}_{+} = \{Q\}$ and
\begin{equation}
 \label{p2:eqn:pythagorean_inequality2}
  \mathscr{I}_{\alpha}(P, P_{\theta}) \ge \mathscr{I}_{\alpha}(P, Q) + \mathscr{I}_{\alpha}(Q, P_{\theta})
\end{equation}
for every $P \in \mathbb{L}$ and every $P_{\theta} \in \hat{\mathbb{M}}^{(\alpha)}_{+}$.

 \item[(b)] If $Q \in \text{cl}(\mathbb{M}^{(\alpha)})$, then $\mathbb{L} \cap \text{cl}(\mathbb{M}^{(\alpha)}) = \{Q\}$ and (\ref{p2:eqn:pythagorean_inequality2}) holds with equality for every $P \in \mathbb{L}$ and every $P_{\theta} \in \text{cl}(\mathbb{M}^{(\alpha)})$.

 \item[(c)] If $Q \in \mathbb{M}^{(\alpha)}$, then $\mathbb{L} \cap \mathbb{M}^{(\alpha)} = \{Q\}$ and (\ref{p2:eqn:pythagorean_inequality2}) holds with equality for every $P \in \mathbb{L}$ and every $P_{\theta} \in \mathbb{M}^{(\alpha)}$.
\end{itemize}
\end{theorem}
\vspace{0.1in}

\begin{IEEEproof}
 (a) By virtue of Theorem \ref{p2:thm:forwardprojection_alpha>1}-(b), we have $Q\in \mathbb{L} \cap \hat{\mathbb{M}}^{(\alpha)}_{+}$. Furthermore, by Theorem \ref{p2:thm:forwardprojection_alpha>1_converse}, any member of $\mathbb{L} \cap \hat{\mathbb{M}}^{(\alpha)}_{+}$ is a forward $\mathscr{I}_{\alpha}$-projection of $R$ on $\mathbb{L}$. Since the forward projection is unique, $\mathbb{L} \cap \hat{\mathbb{M}}^{(\alpha)}_{+}$ must be the singleton $\{Q\}$.

 Let $P_{\theta} \in \hat{\mathbb{M}}^{(\alpha)}_{+}$. We claim that $P_{\theta}$ has $P_{\theta^*} = Q$ as its forward projection on $\mathbb{L}$. Assuming the claim, by Theorem \ref{p2:thm:forwardprojection_alpha>1}-(c), inequality (\ref{p2:eqn:pythagorean_inequality2}) holds.

Let us now proceed to show the claim. By Theorem \ref{p2:thm:forwardprojection_alpha>1_converse}, it suffices to verify that $P_{\theta^*}$ can be written as
\begin{equation}
\label{p2:theta*_in_terms_of_theta1}
 \tilde Z(\tilde\theta)^{\alpha -1} P_{\theta^*}(x)^{\alpha - 1} = \Big[P_{\theta}(x)^{\alpha -1} + (1-\alpha)\sum\limits_{i=1}^k \tilde{\theta}_i f_i(x)\Big]_{+} \quad \forall x
\end{equation}
for some $\tilde Z(\tilde\theta)$ and $\tilde \theta = (\tilde \theta_1, \dots, \tilde \theta_k)$. To see this, by definition of $P_{\theta}$, we have
\begin{equation}
\label{p2:theta_in_terms_of_R}
 Z(\theta)^{\alpha -1} P_{\theta}(x)^{\alpha - 1} = \Big[R(x)^{\alpha -1} + (1-\alpha)\sum\limits_{i=1}^k \theta_i f_i(x)\Big]_{+} \quad \forall x,
\end{equation}
and, by Theorem \ref{p2:thm:forwardprojection_alpha>1}-(b), we have
\begin{equation}
\label{p2:theta*_in_terms_of_R}
 Z(\theta^*)^{\alpha -1} P_{\theta^*}(x)^{\alpha - 1} = \Big[R(x)^{\alpha -1} + (1-\alpha)\sum\limits_{i=1}^k \theta_i^* f_i(x)\Big]_{+} \quad \forall x.
\end{equation}
 Let $x \in \text{Supp}(P_{\theta^*})$. By Definition \ref{p2:defn:m-alphahat}-(a), $x \in \text{Supp}(P_{\theta})$ as well. Hence, we can remove the $[\cdot]_{+}$ operation in (\ref{p2:theta_in_terms_of_R}) and (\ref{p2:theta*_in_terms_of_R}) to get
\[
 Z(\theta)^{\alpha -1} P_{\theta}(x)^{\alpha - 1} = R(x)^{\alpha -1} + (1-\alpha)\sum\limits_{i=1}^k \theta_i f_i(x).
\]
\[
 Z(\theta^*)^{\alpha -1} P_{\theta^*}(x)^{\alpha - 1} = R(x)^{\alpha -1} + (1-\alpha)\sum\limits_{i=1}^k \theta_i^* f_i(x),
\]
Eliminating $R(x)^{\alpha-1}$ from the preceding equations, we get
\[
Z(\theta^*)^{\alpha -1} P_{\theta^*}(x)^{\alpha - 1}  = Z(\theta)^{\alpha -1} P_{\theta}(x)^{\alpha - 1} + (1-\alpha)\sum\limits_{i=1}^k (\theta_i^* - \theta_i) f_i(x),
\]
equivalently,
\begin{eqnarray}
\label{p2:theta*_in_terms_of_theta2}
  \Bigg(\frac{Z(\theta^*)}{Z(\theta)}\Bigg)^{\alpha -1} P_{\theta^*}(x)^{\alpha - 1} = P_{\theta}(x)^{\alpha - 1} + (1-\alpha)\sum\limits_{i=1}^k \frac{(\theta_i^* - \theta_i)}{Z(\theta)^{\alpha -1}} f_i(x).
\end{eqnarray}
This suggests that $\tilde Z(\tilde\theta) = Z(\theta^*)/Z(\theta)$ and $\tilde \theta_i = (\theta_i^* - \theta_i)/Z(\theta)^{\alpha -1}$ should work. Let us now verify that they do, that is, that (\ref{p2:theta*_in_terms_of_theta1}) holds for all $x$ with these choices of $\tilde Z$ and $\tilde \theta$.

The foregoing shows (\ref{p2:theta*_in_terms_of_theta1}) holds for all $x\in \text{Supp}(P_{\theta^*})$. Next, let $x \in \text{Supp}(P_{\theta})\setminus \text{Supp}(P_{\theta^*})$. The right-hand side of (\ref{p2:theta*_in_terms_of_theta2}), upon substitution of (\ref{p2:theta_in_terms_of_R}) without the $[\cdot]_{+}$ operation, becomes
\begin{eqnarray*}
 \frac{R(x)^{\alpha -1} + (1-\alpha)\sum\limits_{i=1}^k \theta_i f_i(x)}{Z(\theta)^{\alpha -1}} + (1-\alpha)\sum\limits_{i=1}^k \frac{(\theta_i^* - \theta_i)}{Z(\theta)^{\alpha -1}} f_i(x)
& = & \frac{R(x)^{\alpha -1} + (1-\alpha)\sum\limits_{i=1}^k \theta_i^* f_i(x)}{Z(\theta)^{\alpha -1}}\\
& \le & 0,
\end{eqnarray*}
as is required for $x \notin \text{Supp}(P_{\theta^*})$. Hence (\ref{p2:theta*_in_terms_of_theta1}) holds for $x \in \text{Supp}(P_{\theta})\setminus \text{Supp}(P_{\theta^*})$ as well, and therefore for all $x\in \text{Supp}(P_{\theta})$. 

Finally, when $x \notin \text{Supp}(P_{\theta})$,
\[
  R(x)^{\alpha -1} + (1-\alpha)\sum\limits_{i=1}^k \theta_i f_i(x) \le 0.
\]
 The right-hand side of (\ref{p2:theta*_in_terms_of_theta2}) then satisfies 
\[
 (1-\alpha)\sum\limits_{i=1}^k \frac{(\theta_i^* - \theta_i)}{Z(\theta)^{\alpha -1}} f_i(x) \le 0
\]
because of condition (b) in Definition \ref{p2:defn:m-alphahat} and $\alpha >1$. This establishes that $P_{\theta^*}$ is of the form (\ref{p2:theta*_in_terms_of_theta1}), and is therefore the forward $\mathscr{I}_{\alpha}$-projection of $P_{\theta}$ on $\mathbb{L}$.

Proofs of (b) and (c) are the same as in $\alpha <1$ case considered in Theorem \ref{p2:thm:orthogonal_intersection_alpha<1}.
\end{IEEEproof}

\vspace{0.1in}

Having established the orthogonality between a linear family and its associated $\alpha$-power-law family, let us now return to the problem of robust estimation discussed in section \ref{p2:subsec:reverse-I-alpha_g1}. As in the case of $\alpha <1$, we show a connection between the MMPLE on the extended $\alpha$-power-law family $\hat{\mathbb{M}}^{(\alpha)}_{+}$, which is a reverse $\mathscr{I}_{\alpha}$-projection on $\hat{\mathbb{M}}^{(\alpha)}_{+}$, and the forward $\mathscr{I}_{\alpha}$-projection on the related linear family.

\vspace{0.1in}

\begin{theorem}
\label{p2:thm:mmple}
 Let $\alpha >1$. Let $\hat{P}$ be a probability measure on $\mathbb{X}$. Let $\mathbb{M}^{(\alpha)}$ be characterized by the probability measure $R$ and the functions $f_i, i=1, \dots, k$. Let $R$ have full support. Let $\mathbb{L}$ be the associated linear family characterized by $f_i, i=1, \dots, k$, and assume that it is nonempty. Define $\tilde{\mathbb{L}}$ as in (\ref{p2:eqn:elltilde}) using $\tilde f_i$ and $\tau_i^R$ as defined in (\ref{p2:eqn:eftilde}) and (\ref{p2:eqn:tau}), respectively. Let $Q$ be the forward $\mathscr{I}_{\alpha}$-projection of $R$ on $\tilde{\mathbb{L}}$. Then the following hold.
\begin{itemize}
 \item[(a)] If $Q \in \mathbb{M}^{(\alpha)}$, then $Q$ is the unique reverse $\mathscr{I}_{\alpha}$-projection of $\hat{P}$ on $\mathbb{M}^{(\alpha)}$.

 \item[(b)] If $Q \in \text{cl}(\mathbb{M}^{(\alpha)})\setminus \mathbb{M}^{(\alpha)}$, then $\hat{P}$ does not have a reverse $\mathscr{I}_{\alpha}$-projection on $\mathbb{M}^{(\alpha)}$. However, $Q$ is the unique reverse $\mathscr{I}_{\alpha}$-projection of $\hat{P}$ on $\text{cl}(\mathbb{M}^{(\alpha)})$.

 \item[(c)] If $Q \notin \text{cl}(\mathbb{M}^{(\alpha)})$, then
\begin{itemize}
 \item[(i)] $\hat{P}$ does not have a reverse $\mathscr{I}_{\alpha}$-projection on $\mathbb{M}^{(\alpha)}$.

 \item[(ii)] $\mathbb{M}^{(\alpha)}$ can be extended to $\hat{\mathbb{M}}^{(\alpha)}_{+}(R, \tilde f_1, \dots, \tilde f_k)$, and $Q$ is the unique reverse $\mathscr{I}_{\alpha}$-projection of $\hat{P}$ on $\hat{\mathbb{M}}^{(\alpha)}_{+}(R, \tilde f_1, \dots, \tilde f_k)$.
\end{itemize}
\end{itemize}

\end{theorem}

\vspace{0.1in}

\begin{IEEEproof}
Only (c)-(i) needs a proof. Proofs of all others follow the same arguments in the proof of Theorem \ref{p2:thm:compression_problem}, but now one uses Theorem \ref{p2:thm:orthogonal_intersection_alpha>1} instead of Corollary \ref{p2:cor:orthogonal_intersection_alpha<1}. 

Let us now prove (c)-(i) by contradiction. Suppose $\hat{P}$ has a reverse $\mathscr{I}_{\alpha}$-projection on $\mathbb{M}^{(\alpha)}$. Call it $P_{\theta^*}$. Since $P_{\theta^*}$ has full support, there is a neighborhood $N$ of $\theta^*$ such that $\theta \in N$ implies $P_{\theta}\in \mathbb{M}^{(\alpha)}$. The first order optimality condition applies, namely
\[
  \frac{\partial}{\partial\theta_i}\mathscr{I}_{\alpha}(\hat{P},P_{\theta})\bigg |_{\theta = \theta^*} = 0, \, \, i=1, \dots, k.
\]
We claim that this implies
\begin{eqnarray}
 \label{p2:eqn:exp-fi-tilde-zero}
 \sum\limits_x P_{\theta^*}(x) \tilde f_i(x) = 0, \, \,  i=1, \dots, k.
\end{eqnarray}
But then $P_{\theta^*} \in \tilde{\mathbb{L}}$ and so $P_{\theta^*} = Q$, a contradiction to $Q \notin \text{cl}(\mathbb{M}^{(\alpha)})$.

We now proceed to prove the claim (\ref{p2:eqn:exp-fi-tilde-zero}). Observe that, since $P_{\theta}\in \mathbb{M}^{(\alpha)}$, by Definition \ref{p2:defn:alpha-power-family}, we have
\begin{eqnarray}
\label{p2:eqn:theta-intermsof-R}
 Z(\theta)^{\alpha-1} P_{\theta}(x)^{\alpha-1} = R(x)^{\alpha-1} + \sum\limits_j \theta_j \tilde f_j(x),
\end{eqnarray}
and so
\begin{eqnarray}
\label{p2:eqn:p-hatandp-theta}
Z(\theta)^{\alpha-1} \sum\limits_x \hat{P}(x) P_{\theta}(x)^{\alpha-1}
& = & \sum\limits_x \hat{P}(x) R(x)^{\alpha-1} + \sum\limits_j \theta_j \Big(\sum\limits_x \hat{P}(x) \tilde f_j(x)\Big)\nonumber\\
& = & \sum\limits_x \hat{P}(x) R(x)^{\alpha-1},
\end{eqnarray}
where the last equality holds because $\hat{P}\in \tilde{\mathbb{L}}$. Also,
\begin{eqnarray}
\label{p2:eqn:p-thetaandp-theta}
 \sum\limits_x P_{\theta}(x)^{\alpha} & = & \sum\limits_x \Big[P_{\theta}(x)^{\alpha-1}\Big]^{\frac{\alpha}{\alpha-1}}\nonumber\\
& = & Z(\theta)^{-\alpha} \sum\limits_x \Big[ R(x)^{\alpha-1} + \sum\limits_j \theta_j \tilde f_j(x)\Big]^\frac{\alpha}{\alpha-1}.
\end{eqnarray}
Substituting (\ref{p2:eqn:p-hatandp-theta}) and (\ref{p2:eqn:p-thetaandp-theta}) into (\ref{p2:alphadiv_expanded}) and taking the partial derivative, we get
\begin{eqnarray*}
 \frac{\partial}{\partial\theta_i}\mathscr{I}_{\alpha}(\hat{P},P_{\theta})
& = & \frac{\alpha}{1-\alpha}\frac{\partial}{\partial\theta_i} \log Z(\theta)^{1-\alpha} + \frac{\partial}{\partial\theta_i} \log Z(\theta)^{-\alpha} + \frac{\partial}{\partial\theta_i} \log \sum\limits_x \Big[ R(x)^{\alpha-1} + \sum\limits_j \theta_j \tilde f_j(x)\Big]^\frac{\alpha}{\alpha-1}\\
& = & \frac{\partial}{\partial\theta_i} \log \sum\limits_x \Big[ R(x)^{\alpha-1} + \sum\limits_j \theta_j \tilde f_j(x)\Big]^\frac{\alpha}{\alpha-1}\\
& = & \frac{1}{A} \cdot\frac{\alpha}{1-\alpha} \sum\limits_x \Big[ R(x)^{\alpha-1} + \sum\limits_j \theta_j \tilde f_j(x)\Big]^\frac{1}{\alpha-1} \tilde f_i(x)\\
& = & \frac{1}{A} \cdot\frac{\alpha}{1-\alpha} Z(\theta) \sum\limits_x P_{\theta}(x) \tilde f_i(x),
\end{eqnarray*}
 where $A = \sum_x \Big[ R(x)^{\alpha-1} + \sum\limits_j \theta_j \tilde f_j(x)\Big]^\frac{\alpha}{\alpha-1}$, and the last equality follows from (\ref{p2:eqn:theta-intermsof-R}). Thus,
\[
 \frac{\partial}{\partial\theta_i}\mathscr{I}_{\alpha}(\hat{P},P_{\theta})\bigg |_{\theta = \theta^*} = 0 \implies \sum\limits_x P_{\theta^*}(x) \tilde f_i(x) = 0,
\]
thereby proving the claim.
\end{IEEEproof}

\section{Epilogue}
\label{p2:sec:concludingRemarks}
We now provide some concluding remarks. Our focus has primarily been on the geometric relation between the $\alpha$-power-law and the linear families. This geometric relation enabled us to characterize the reverse $\mathscr{I}_{\alpha}$-projection on an $\alpha$-power-law family $\mathbb{M}^{(\alpha)}:=\mathbb{M}^{(\alpha)}(R, f_1, \dots, f_k)$ as a forward $\mathscr{I}_{\alpha}$-projection on a linear family. The procedure is as follows. 

\vspace{0.15in}

\begin{quote}
 ``Given the family $\mathbb{M}^{(\alpha)}$, sweep through a collection of linear families (\ref{p2:eqn:elltilde})-(\ref{p2:eqn:tau}) orthogonal to \ $\mathbb{M}^{(\alpha)}$ by varying $\tau_i^R, i=1, \dots, k$, and find the linear family $\tilde{\mathbb{L}}$ that contains $\hat{P}$. Then find the forward $\mathscr{I}_{\alpha}$-projection of $R$ on $\tilde{\mathbb{L}}$; call it $Q$. If $Q\in \mathbb{M}^{(\alpha)}$, then $Q$ is the reverse $\Ia$-projection of $\hat{P}$ on the $\mathbb{M}^{(\alpha)}$. If $Q\in \text{cl}(\mathbb{M}^{(\alpha)})\setminus \mathbb{M}^{(\alpha)}$, then $\hat{P}$ does not have a reverse $\Ia$-projection on $\mathbb{M}^{(\alpha)}$. But $Q$ attains the minimum in the closure.''
\end{quote}

\vspace{0.15in}

The cases $\alpha <1$ and $\alpha >1$ have different characteristics. The $\alpha <1$ case is similar to $\alpha =1$ and one always has $\tilde{\mathbb{L}}\cap \text{cl}(\mathbb{M}^{(\alpha)}) = \{Q\}$. On the other hand, when $\alpha >1$, it is possible that $\tilde{\mathbb{L}}\cap \text{cl}(\mathbb{M}^{(\alpha)}) = \emptyset$, and $Q\notin \text{cl}(\mathbb{M}^{(\alpha)})$. Then $\hat{P}$ does not have a reverse $\mathscr{I}_{\alpha}$-projection on $\mathbb{M}^{(\alpha)}$. One then needs to extend $\mathbb{M}^{(\alpha)}$ to make it intersect $\tilde{\mathbb{L}}$. We showed that the extension $\hat{\mathbb{M}}^{(\alpha)}_{+}$ is just right and satisfies $\tilde{\mathbb{L}}\cap \hat{\mathbb{M}}^{(\alpha)}_{+} = \{Q\}$. However, $Q$, in the intersection $\tilde{\mathbb{L}}\cap \hat{\mathbb{M}}^{(\alpha)}_{+}$, is no longer the reverse $\mathscr{I}_{\alpha}$-projection of $\hat{P}$ on $\text{cl}(\mathbb{M}^{(\alpha)})$. It would be interesting to see if $Q$ can be used to simplify the computation of the true reverse $\mathscr{I}_{\alpha}$-projection of $\hat{P}$ on $\text{cl}(\mathbb{M}^{(\alpha)})$.

Our characterization has algorithmic benefits since the forward $\mathscr{I}_{\alpha}$-projection is a minimization of a quasiconvex function subject to linear constraints. Standard techniques are available to solve such problems, for example, via a sequence of convex feasibility problems \cite[Sec.~4.2.5]{2004xxCOP_BoyVan}, or via a sequence of simpler forward projections on single-constraint linear families \cite[Th. 16, Rem. 13]{2014xxManuscript1_KumSun}.

\appendices

\section{Weak dependence of the $\alpha$-power-law family on $R$}

The following result shows that the $\alpha$-power-law family depends on $R$ only in a weak manner, and that any member of $\mathbb{M}^{(\alpha)}$ could equally well play the role of $R$. The same result is well-known for an exponential family.

\vspace*{.1in}
\begin{proposition}
\label{p2:app:weakdependence}
If $\alpha >1$, let $R$ have full support. Consider the $\mathbb{M}^{(\alpha)}(R,f_1,\dots,f_k)$ as in Definition \ref{p2:defn:alpha-power-family}. Fix $P_{\theta^*}\in \mathbb{M}^{(\alpha)}(R,f_1,\dots,f_k)$. Then $\mathbb{M}^{(\alpha)}(P_{\theta^*},f_1,\dots,f_k) = \mathbb{M}^{(\alpha)}(R,f_1,\dots,f_k)$.
\end{proposition}
\vspace*{.1in}


\begin{IEEEproof}
Write $\mathbb{M}^{(\alpha)}$ for $\mathbb{M}^{(\alpha)}(R,f_1,\dots,f_k)$ and $\tilde{\mathbb{M}}^{(\alpha)}$ for $\mathbb{M}^{(\alpha)}(P_{\theta^*},f_1,\dots,f_k)$. We will check that an arbitrary element $P_{\theta}\in \mathbb{M}^{(\alpha)}$ is an element of $\tilde{\mathbb{M}}^{(\alpha)}$. This will establish $\mathbb{M}^{(\alpha)}\subset \tilde{\mathbb{M}}^{(\alpha)}$. The converse holds by symmetry.

From the formula for $P_{\theta^*}$, observe that
\[
  P_{\theta^*}(x)^{\alpha-1} = Z(\theta^*)^{1-\alpha} \Big[ R(x)^{\alpha-1} + (1-\alpha) \sum\limits_{i=1}^k \theta_i^* f_i(x) \Big] \quad \forall x,
\]
and so
\begin{equation}
  \label{p2:eqn:S-P-translation}
  R(x)^{\alpha-1} = Z(\theta^*)^{\alpha-1} P_{\theta^*}(x)^{\alpha-1} - (1-\alpha) \sum\limits_{i=1}^k \theta_i^* f_i(x) \quad \forall x.
\end{equation}
Substitute this into the formula for $P_{\theta}$ in (\ref{p2:eqn:power_law_family}) to get
\begin{eqnarray*}
  P_{\theta}(x)^{\alpha-1} & = & Z(\theta)^{1-\alpha} \Big[ Z(\theta^{*})^{\alpha-1} P_{\theta^*}(x)^{\alpha-1} - (1-\alpha) \sum\limits_{i=1}^k \theta_i^* f_i(x) + (1-\alpha) \sum\limits_{i=1}^k \theta_i f_i(x) \Big] \\
    & = & \left(\frac{Z(\theta^*)}{Z(\theta)}\right)^{\alpha-1} \Big[ P_{\theta^*}(x)^{\alpha-1} + (1-\alpha) \sum\limits_{i=1}^k \frac{\theta_i - \theta_i^*}{Z(\theta^*)^{\alpha-1}} f_i(x)  \Big] \\
    & = & \tilde{Z}(\xi)^{1-\alpha} \Big[ P_{\theta^*}(x)^{\alpha-1} + (1-\alpha) \sum\limits_{i=1}^k \xi_i f_i(x) \Big],
\end{eqnarray*}
where $\xi = (\theta - \theta^*)/Z(\theta^*)^{\alpha-1}$, and $\tilde{Z}(\xi) = Z(\theta) / Z(\theta^*)$. Thus, $P_{\theta}\in \tilde{\mathbb{M}}^{(\alpha)}$.
\end{IEEEproof}

\vspace*{.1in}

Change of reference from $R$ to $P_{\theta^*}$ merely amounts to a translation and rescaling of the parameter space.

\section*{Acknowledgements}

We thank the reviewers whose comments/suggestions helped improve this manuscript enormously.

\bibliographystyle{IEEEtran}
{
\bibliography{reverse-projection-26APR15-revision}
}

\end{document}